\documentclass[lettersize,journal]{IEEEtran}
\usepackage{amsmath,amsfonts}
\usepackage{algorithmic}
\usepackage{algorithm}
\usepackage{array,makecell,bbding,pifont}
\usepackage[caption=false,font=normalsize,labelfont=sf,textfont=sf]{subfig}
\usepackage{textcomp}
\usepackage{stfloats, threeparttable}
\usepackage{url}
\usepackage{verbatim}
\usepackage{graphicx}
\usepackage{cite}
\usepackage{url}
\usepackage{colortbl}
\usepackage{xcolor}
\usepackage{array}
\usepackage{color}
\usepackage{hyperref}
\usepackage{upgreek}
\usepackage{enumerate}
\usepackage[figuresright]{rotating}
\usepackage{bbding}

\begin{document}
	
	\title{Generative AI-Driven Human Digital Twin in IoT-Healthcare: A Comprehensive Survey}
	
	\author{Jiayuan Chen, You Shi, Changyan Yi,~\IEEEmembership{Member,~IEEE}, Hongyang Du, Jiawen Kang, Dusit Niyato,~\IEEEmembership{Fellow,~IEEE}

		\IEEEcompsocitemizethanks{\IEEEcompsocthanksitem Copyright (c) 20xx IEEE. Personal use of this material is permitted. However, permission to use this material for any other purposes must be obtained from the IEEE by sending a request to pubs-permissions@ieee.org. \IEEEcompsocthanksitem J. Chen Y. Shi and C. Yi are with the College of Computer Science and Technology, Nanjing University of Aeronautics and Astronautics, Nanjing, Jiangsu, 211106, China. (E-mail: \{jiayuan.chen, shyou changyan.yi\}@nuaa.edu.cn).
						\IEEEcompsocthanksitem
			H. Du and D. Niyato are with the School of Computer Science and Engineering, Nanyang Technological University, Singapore. (Email: hongyang001@e.ntu.edu.sg and dniyato@ntu.edu.sg).
							\IEEEcompsocthanksitem
			J. Kang is with the School of Automation, Guangdong University of Technology, China. (Email: kjwx886@163.com).
			\\
		}
		
	}

	\maketitle
	
	\begin{abstract}
	The Internet of things (IoT) can significantly enhance the quality of human life, specifically in healthcare, attracting extensive attentions to IoT-healthcare services. Meanwhile, the human digital twin (HDT) is proposed as an innovative paradigm that can comprehensively characterize the replication of the individual human body in the digital world and reflect its physical status in real time. Naturally, HDT is envisioned to empower IoT-healthcare beyond the application of healthcare monitoring by acting as a versatile and vivid human digital testbed, simulating the outcomes and guiding the practical treatments. However, successfully establishing HDT requires high-fidelity virtual modeling and strong information interactions but possibly with scarce, biased and noisy data. Fortunately, a recent popular technology called generative artificial intelligence (GAI) may be a promising solution because it can leverage advanced AI algorithms to automatically create, manipulate, and modify valuable while diverse data. This survey particularly focuses on the implementation of GAI-driven HDT in IoT-healthcare. We start by introducing the background of IoT-healthcare and the potential of GAI-driven HDT. Then, we delve into the fundamental techniques and present the overall framework of GAI-driven HDT. After that, we explore the realization of GAI-driven HDT in detail, including GAI-enabled data acquisition, communication, data management, digital modeling, and data analysis. Besides, we discuss typical IoT-healthcare applications that can be revolutionized by GAI-driven HDT, namely personalized health monitoring and diagnosis, personalized prescription, and personalized rehabilitation. Finally, we conclude this survey by highlighting some future research directions.

	\end{abstract}
	
	
	\section{Introduction}\label{SE1}
	\subsection{Background}
		\IEEEPARstart{T}{he} Internet of the Things (IoT) refers to a group of Internet-connected devices that have computing capability, are easy to identify, and are able to share data over the Internet without the need for direct touch \cite{415}. The IoT progression has led to unmatched connectivity among devices, paving the way for the creation of intelligent environments, enabling new ways of working, communicating, interacting, entertaining, and living \cite{416}.  Notably, IoT is leading to a paradigm shift in the healthcare industry, which are termed IoT-healthcare \cite{349, 350}. A global healthcare organization estimates that roughly 17.5 million lives are lost annually due to inefficiencies in collecting and analyzing health data \cite{356}. Fortunately, the advancement of IoT-healthcare presents a promising avenue to tackle these challenges \cite{352}. By enabling the real-time collection and transmission of pertinent health data to servers for in-depth healthcare analysis via personal IoT devices, it can prompt timely healthcare alerts and proactive interventions for saving lives \cite{351, 418}. IoT-healthcare has been widely applied in healthcare monitoring, including blood glucose monitoring, cardiac monitoring, respiration monitoring and blood pressure monitoring, providing people with pervasive healthcare services \cite{352, 351}.

Meanwhile, digital twin (DT) has attracted substantial attentions since 2002 when Michael Grieves delivered a presentation at the University of Michigan \cite{420}. With the continuous progression, DT, nowadays, is characterized as a smart, evolving system that precisely mirrors a physical entity at various granularity levels, and monitors, controls, and optimizes the physical objects during its lifecycle \cite{419}. When adopting DT in the human-centric systems, a revolutionary technology known as Human Digital Twin (HDT) has emerged. HDT is a promising technology and game changer for IoT-healthcare, taking this field to another level. HDT can create a digital replica of a human body, comprehensively and precisely characterizing each individual in the digital space while reflecting its physical status in real time \cite{054, 3,4,5,6}.

	Besides, with visualization and interaction characteristics, HDT is envisioned as a versatile and vivid human digital testbed for revolutionizing the IoT-healthcare, beyond healthcare monitoring applications mentioned above. For instance, a patient's HDT can be implemented in silico treatment simulations and experiments, facilitating the development of finely-tailored, personalized treatment plans \cite{18}. Additionally, a doctor's HDT with expert-level medical knowledge can be a personal 24/7 doctor to answer the patients' queries \cite{235}.

	The successful establishment and implementation of HDT largely depend on the high-fidelity human modeling, supported by comprehensive individual-level data encompassing appearance, movement, and physiological data, acquired from multi-source, such as IoT devices.
	In addition, as an intelligent human digital testbed, HDT needs to generate various human-like feedback during immersive real-virtual interactions. These include, for example, providing intuitive feedback on drug-disease responses, and simulating haptic feedback to replicate the tactile sensations experienced by humans in real-world scenarios. All such requirements are difficult to meet due to several crucial reasons, e.g., data scarcity, bias, noise and intricate digital modeling. Fortunately, generative artificial intelligence (GAI) has been recognized as a promising technology that can effectively fulfill or assist the implementation of HDT for IoT-healthcare \cite{8}.

	GAI can leverage advanced AI algorithms to automatically create, manipulate, and modify valuable while diverse data \cite{du2023beyond,10}. Specifically, GAI models, such as generative adversarial network (GAN), variational autoencoder (VAE), transformer, and diffusion model, with their powerful creativities and data analysis abilities can generate ultra-realistic individual-level data and make informed decisions for HDT in IoT-healthcare, which will be elaborated in Section \ref{GAIforHDT}. Thus, our survey focuses on how GAI enables HDT in IoT-healthcare, namely GAI-driven HDT in IoT-healthcare.
	
\subsection{Review Methodology} \label{method}
We identify three specific research questions that lead to the accomplishment of this work in the following.

1) RQ1: What is the definition of HDT? And what GAI models can be potentially applied in HDT?

2) RQ2: How GAI can enable the implementation of HDT?

3) RQ3: How GAI-driven HDT empowers IoT-healthcare? 

We perform this survey in a systematic way following the guidelines, preferred reporting items for systematic reviews and meta-analyses (PRISMA) \cite{421}, which consists of three phases, i.e., identification, screening, eligibility and included.

\emph{1. Identification:} For the literature review, we used three scientific platforms: Google Scholar, Scopus and Web of Science. The search string is defined as shown in Table \ref{string}. Only papers written in English were included. Using these criteria, we found 2,416 papers on Google Scholar, 900 on Scopus, and 1,679 on Web of Science. We then eliminated duplicates from all the papers found. Additionally, some papers that were not highly relevant, potentially due to certain platforms ignoring logical operators (i.e., AND and OR), were also filtered out. The total number of selected papers in this phase was 1,736.

\emph{2. Screening:} In this phase, we evaluated the selected papers in identification phase on the basis of their titles and abstracts. In this survey, we focus on the application of GAI techniques in building HDT from the perspective of IoT-healthcare. Specifically, GAI techniques are expected to enable each component of HDT, including data acquisition, communication, data management, digital modeling, and data analysis, which will be elaborated in Section \ref{GAIforHDT}. All of these GAI-driven components are required to potentially empower IoT-healthcare. Thus, following these criteria, we further filtered these 1736 papers to be 369 papers by examining their titles and abstracts.

\emph{3. Eligibility and Included:} Since the titles and abstracts of some papers were not clear enough to be directly evaluated, a full-text screening was performed on these 369 papers during this phase. We followed the criteria from the screening phase. Finally, a total of 172 papers were included in this paper.

\begin{table}[!t]
	\centering
	\caption{Search String of Acquiring the Literature.}\label{string}
	\includegraphics[width=0.94\columnwidth]{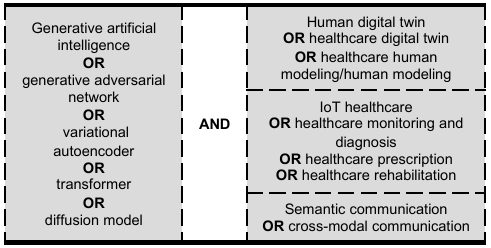}
\end{table}

\subsection{Related Work and Contributions}
Given the increasing interest of GAI-driven HDT in IoT-healthcare, several surveys and tutorials have been recently published \cite{s1, s2, s3, s4, s5, s6}. Table \ref{related} presents a comparison of these related works compared with ours.
\begin{table*}[htbp]
	\centering
	\caption{Comparison of The Related Work with Our Survey.}\label{related}
	\includegraphics[width=0.94\textwidth]{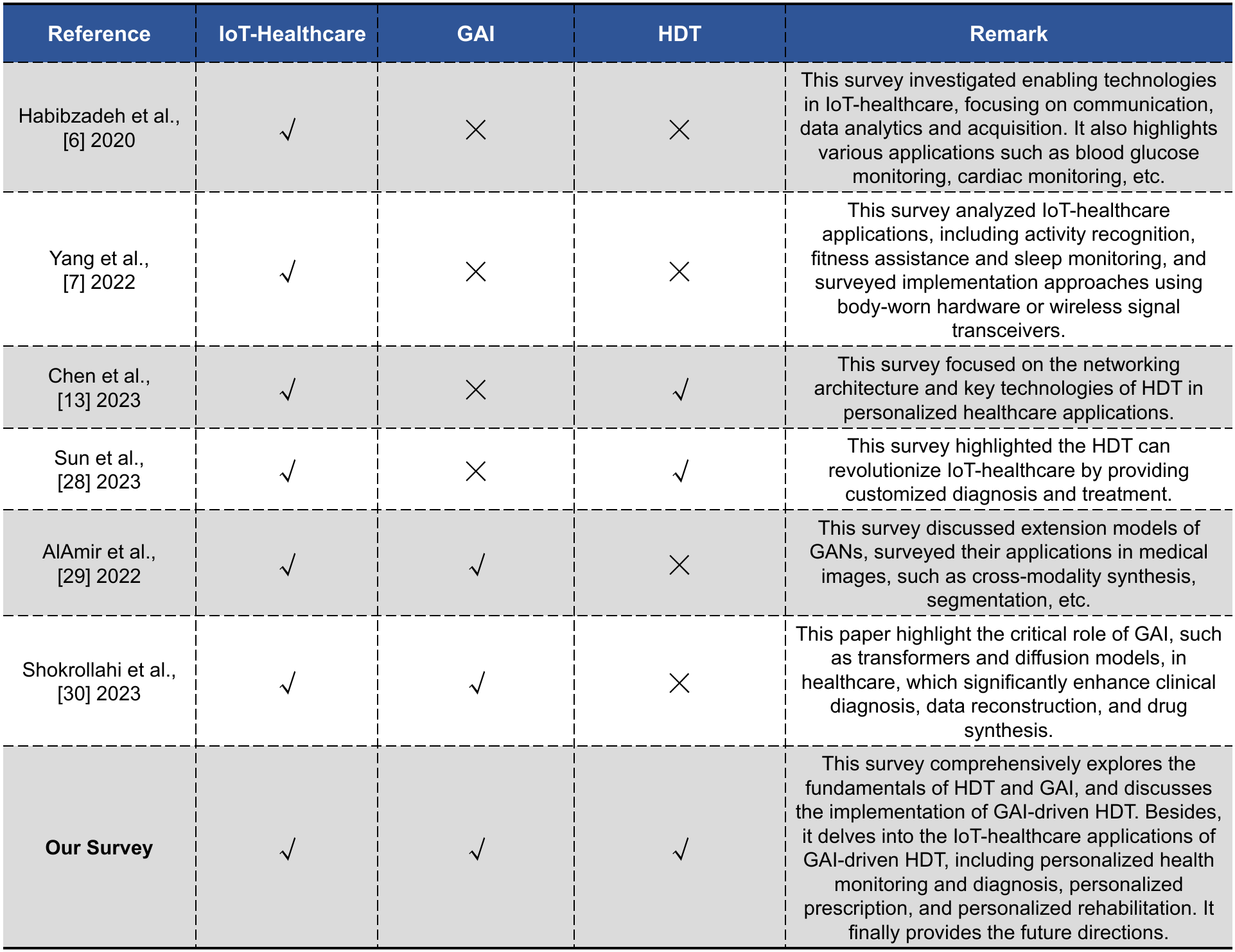}
\end{table*}

Specifically,
Habibzadeh et al. in \cite{352} surveyed the existing and emerging technologies that can enable IoT-healthcare. It presented the enabling technologies by investigating three of IoT-healthcare primary components: 1) sensing and data acquisition; 2) communication; and 3) data analytics and inference. Based on these, they also highlight some IoT-healthcare applications, including blood glucose monitoring, cardiac monitoring, respiration monitoring, blood pressure monitoring, among others. Yang et al. in \cite{351} investigated the IoT-healthcare applications with high relevance to daily health routines, including activity recognition, fitness assistance, vital signs monitoring, daily dietary tracking, and sleep monitoring. Additionally, they surveyed the ways of implementing these applications based on leveraging of sensors, such as device-based paradigms using hardware on the body, and device-free paradigms using wireless signal transceivers. However, these surveys mainly focused on healthcare monitoring enabled by IoT, ignoring the power of integration of HDT, which can significantly enrich the application of IoT-healthcare. 

With the prevalence of HDT, there are several surveys are liberating the power of HDT in IoT-healthcare. Chen et al. in \cite{4} comprehensively explored the networking architecture and key supporting technologies for realizing HDT in personalized healthcare. Specifically, the networking architecture consisted of data acquisition, communication, computation, data management, data analysis and decision making layers. They surveyed the enabling technologies for each layer. Additionally, they delved into the application of HDT in personalized healthcare, including personalized diagnosis, prescription, surgery, and rehabilitation. Sun et al. in \cite{353} highlighted that the HDT can revolutionize IoT-healthcare by providing customized diagnosis and treatment. They revealed that by using a patient's HDT, the medical system can predict the patient's immune response to infection or injury, which can help doctors diagnose diseases precisely. Additionally, they investigated that the patient's HDT can be used as a vivid digital testbed before the prescription or surgery, thereby supporting personalized treatment in a non-invasive manner. Furthermore, they surveyed more specific applications of HDT in IoT-healthcare, including cardiovascular disease, surgery, pharmacy, orthopaedics and COVID-19. However, none of them discuss the role of GAI for HDT in IoT-healthcare.

GAI, with its superior data generation and analysis capabilities, has attracted a myriad of researches recently, and many of them focused on its application in healthcare. AlAmir et al. in \cite{354} discussed the recent advancements in GANs, particularly in the healthcare field. Specifically, they investigated the extension models of GANs by classifying and introducing them individually. Then, they surveyed the applications of these GAN models in medical images, including cross-modality synthesis, segmentation, augmentation, detection, classification, registration and reconstruction. Shokrollahi et al. in \cite{355} delved into the critical role of GAI, such as transformers and diffusion models,  in healthcare applications, including medical imaging, protein structure prediction, clinical documentation, diagnostic assistance, radiology interpretation, clinical decision support, drug design and molecular representation. Such applications have significantly enhanced clinical diagnosis, data reconstruction, and drug synthesis.

Unlike the above surveys that broadly cover IoT-healthcare, HDT, and GAI, ours specifically investigates the pivotal role of GAI in HDT within IoT-healthcare by providing comprehensive and up-to-date review of the current works. GAI's exceptional capabilities in data generation and analysis are crucial for improving all aspects of HDT implementation, from data collection and management to digital modeling and analysis. Our work marks the first in-depth survey of GAI-driven HDT in IoT-healthcare, offering insights into its application in personalized health monitoring adiagnosis, prescription, and rehabilitation. We also highlight current challenges and envisage future directions, ensuring our survey captures the emerging trends in this evolving field. The contributions of this survey can be summarized as follows, which also aims to answer to research questions listed in Section \ref{method} :
\begin{itemize}
	\item We thoroughly review the HDT and GAI technique, including differences between HDT and the conventional DT and the framework of HDT, as well as the popular GAI models.
	\item We explore the implementation of GAI-driven HDT. We comprehensively explain how GAI enables each component of HDT's framework, including data acquisition, digital modeling, communication,data management and data analysis.
	\item We survey GAI-driven HDT in IoT-healthcare applications, including personalized health monitoring and diagnosis, prescription, and rehabilitation.
	\item We outline several open issues and future directions in GAI-driven HDT in IoT-healthcare, helping to drive the development of this field.

\end{itemize}

The rest of this paper is organized as follows: Section \ref{SE2} aims to answer RQ1 in Section \ref{method}, which presents the fundamentals of HDT and GAI, and gives an overview of the framework of GAI-driven HDT. Section \ref{GAIforHDT} aims to answer RQ2 in Section \ref{method}, where the implementation of GAI-driven HDT is analyzed in detail. Section \ref{SE4} aims to answer RQ3 in Section \ref{method}, which surveys the IoT-healthcare application of GAI-driven HDT, including personalized health monitoring and diagnosis, personalized prescription, and personalized rehabilitation. Section \ref{SE5} explores several future research directions of GAI-driven HDT in IoT-healthcare. Section \ref{SE6} concludes this survey paper.

\section{Fundamentals of HDT and GAI} \label{SE2}

\subsection{Human Digital Twin} \label{SE2.1}

HDT as the versatile and vivid digital portrayal of individual, is breathing life into the digital world. According to Emergen Research, the global market for HDT will grow from $29.51\$$ billion in 2022 to about $530\$$ billion in 2032 \cite{13}. With the  continuous advancement of HDT, the IoT-healthcare is being revolutionized by HDT \cite{4,5,7, 8}.

HDT, focuses on digital replicas of human beings while the conventional DT limits the attention to non-living physical entities, e.g., machines \cite{22} and networks \cite{21}. Then, several distinguishing characteristics between HDT and the conventional DT are detailed below and summarized in Table \ref{Differe}.
\begin{table*}[htbp]
	\centering
	\caption{Differences Between HDT and the Conventional DT.}\label{Differe}
	\includegraphics[width=0.9\textwidth]{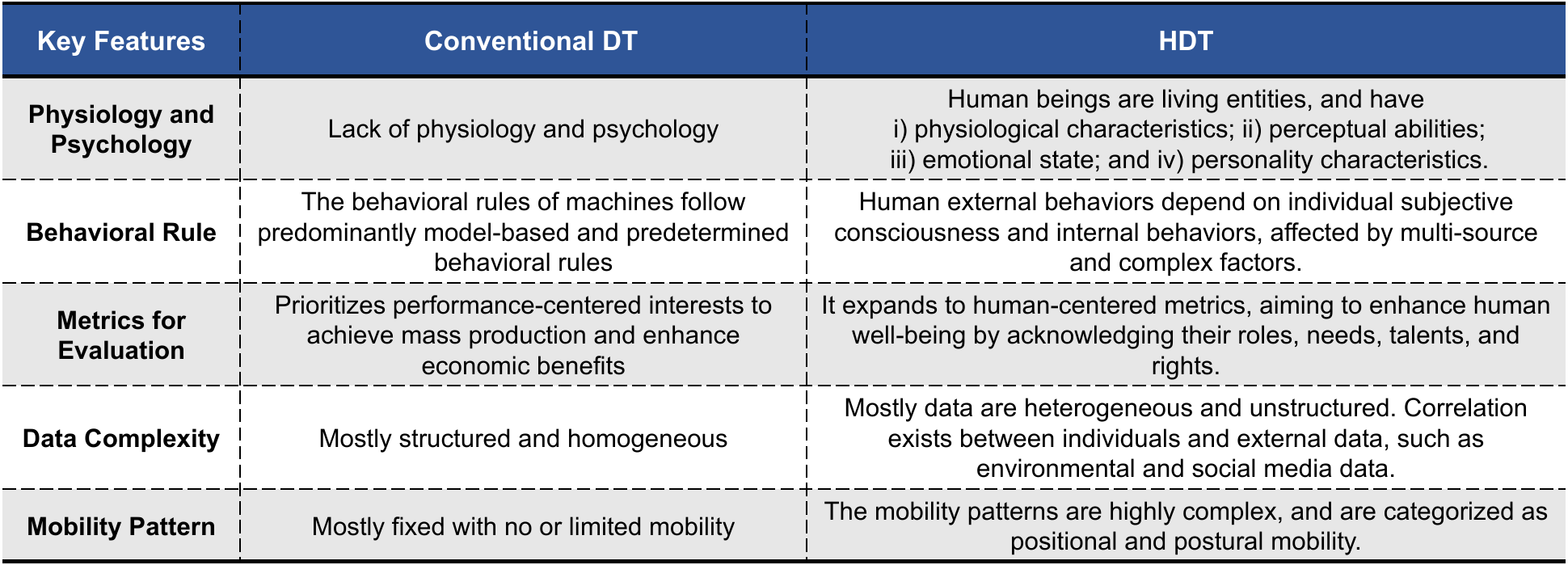}
\end{table*}
\begin{itemize}
	\item Physiology and psychology: The most significant difference between HDT and conventional DT is physiology and psychology \cite{20, 23, 24}. This includes attributes such as: i) physiological characteristics, e.g., brain electrophysiologic signals, blood oxygen level and heart rate; ii) perceptual abilities, e.g., visual sensitivity, pressure sensitivity and temperature sensitivity; iii) emotional state, e.g., happiness, depression and anxiety; iv) personality characteristics, e.g., personality type, propensity to trust, and propensity towards suspicion.
	\item Behavioral rule: Human beings' external behaviors highly depend on the individual subjective consciousness. Particularly, internal behaviors, such as the progression of diseases and emotional states, are generally influenced by multi-source and complex factors, including external environments. In contrast, machines generally follow predominantly model-based and predetermined behavioral rules \cite{25, 26}. Therefore, humans are highly complex systems with greater uncertainty levels than machines. The abstract processes of human beings in HDT are significantly more challenging than those of machines in the conventional DT.
	\item Metrics for evaluation: The human-centered paradigm of HDT requires improving human beings' well-being by considering their roles, needs, talents and rights. Meanwhile, the conventional DT commonly takes the performance-centered interest first for improving production and  economic benefits. Specifically, the conventional DT typically prioritize metrics such as efficiency, productivity, effectiveness, and profitability. However, HDT extends the scope of metrics to contain usability, user experience, etc \cite{3}.
	\item Data complexity: Human beings are more heterogeneous and unstructured than non-living machines. Consequently, unlike the conventional DT,  building a high-fidelity digital representation model of any human entity in HDT requires diverse and complex data from multiple sources. In addition to physiological data, unstructured data from environmental factors and social media play a crucial role in abstracting human virtual twins. This is due to the significant correlation between human beings and such external data sources \cite{4}.
	\item Mobility pattern:  Unlike the position-fixed machines in the conventional DT, the mobility patterns in HDT may be highly predictable. The mobility patterns of human beings can be categorized into human positional and postural mobility. Positional mobility, like a person moving from indoors to outdoors, may cause radio frequency (RF) propagation characteristics to change and even the service migrations. Additionally, the postural mobility of a human, like lying, sitting, walking, may cause signal strength to fluctuate due to the influence of human bodies on the path loss, known as body shadowing \cite{4}.
\end{itemize}

As the distinct features of HDT outlined above, the successful implementation of HDT relies on five essential components: data acquisition, communication, data management, digital modeling, and data analysis \cite{3,4,5,27}. These components form the framework of HDT. Note that, in HDT, the physical entity, i.e., the individual, in the physical world is called a physical twin (PT), while the corresponding virtual one in the digital world is called a virtual twin (VT). Section \ref{GAIforHDT} delves into how GAI facilitates the actualization of HDT by bolstering each component within the framework of HDT. Before this, we introduce the GAI technique, unveiling its potential for enabling HDT.

\subsection{GAI Techniques}
In this subsection, we delve into the recent primary trends of GAI models, including generative adversarial network, variational autoencoder, transformer and diffusion model \cite{du2023beyond,11}. These models find frequent applications in HDT for IoT-healthcare \cite{429, 430, 431, 83}.

\begin{figure}[!t]
	\centering
	\includegraphics[width=0.9\columnwidth]{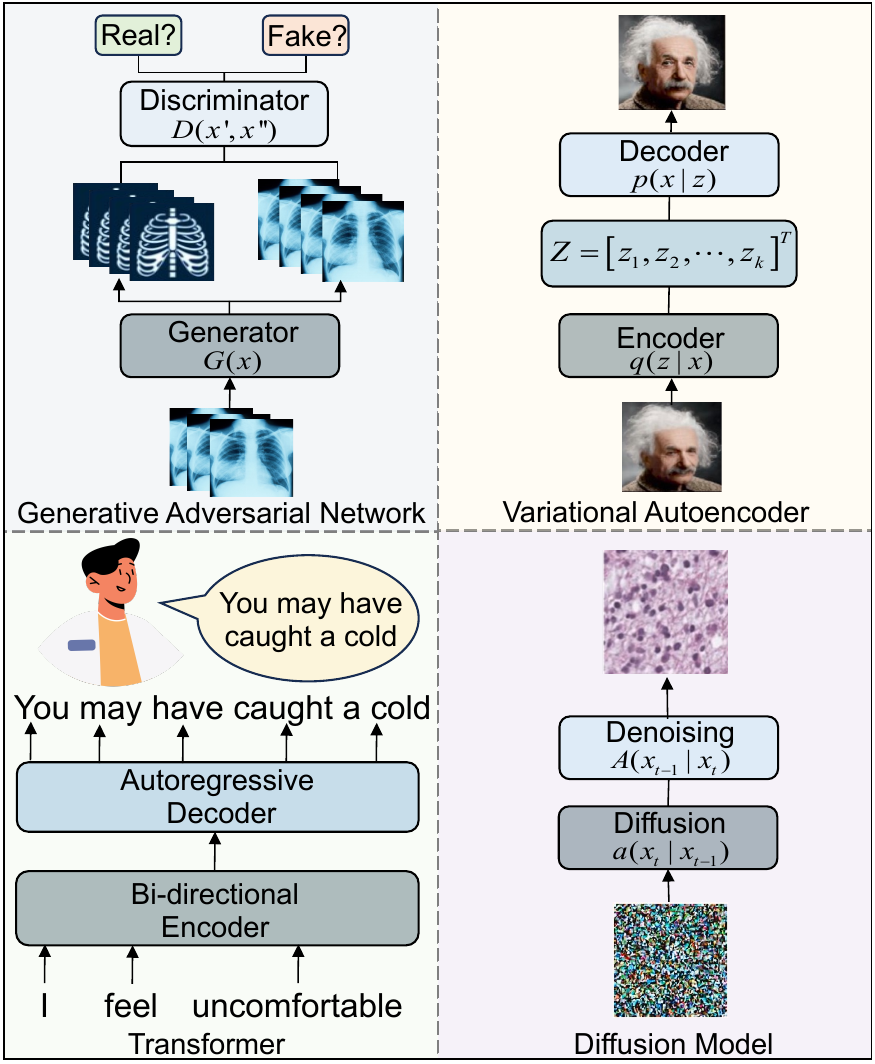} \\
	\caption{The workflow of recent primary trends of GAI models, including generative adversarial network, variational autoencoder, transformer and diffusion model.}\label{Gen}
\end{figure}

\emph{Generative adversarial network}:
As depicted in Fig. \ref{Gen}, the GAN comprises two neural networks engaged in a competitive process to create new samples resembling a specific distribution \textcolor{blue}{\cite{035}}. The first network, the generator, aims to produce synthetic samples by comprehending the underlying distribution of the training data. Meanwhile, the discriminator, the second network, distinguishes between real and synthetic data generated by the generator. Its task is to accurately differentiate real samples and provide feedback to enhance the quality of the generated samples. Throughout the GAN training process, these two networks iteratively refine their performance adversarially, engaging in a competitive interplay until they achieve a stable equilibrium. This continuous refinement allows the generator to create more realistic samples while enabling the discriminator to better distinguish between real and synthetic data.

GAN is significant for HDT in IoT-healthcare, because of its ability to enrich training data, by generating realistic synthetic data. It can ensure patient privacy through anonymized data creation, facilitate data imputation with their cross-modality data translation capabilities, and aid in early anomaly detection for health conditions \cite{429}. For example, a heart dynamics, represented by a ordinary differential equations (ODE) system, was integrated into the training of a GAN to generate biologically plausible electrocardiogram (ECG) data \cite{313}, which can be used to solve the ECG data scarcity issue when digital modeling of cardiac activities. Additionally, a chained GAN-based approach, which connected multiple GAN models, was proposed to simulate the pathology of tissues in the human body. Specifically, it included three steps: i) shape generation with a mask generator; ii) initial simulation via domain-specific heuristics; iii) detail refinement with a refining generator. This approach can be used to model tissue digital twin in HDT, as well as help diagnose diseases and predict their progress \cite{286}.

\emph{Variational autoencoder}:
The core idea of VAE is to transform input data to a low-dimensional latent space representation \cite{036}. Illustrated in Fig. \ref{Gen}, VAE comprises two neural networks. The first network, known as the encoder, maps input data to a latent space, often assumed to follow a Gaussian distribution characterized by learned mean and variance parameters. In contrast, the other network, the decoder, undertakes the task of reconstructing the original input data from a sample drawn from the latent space distribution. The decoder aims to generate a reconstructed sample closely resembling the input data. Throughout the training process, the encoder and decoder parameters are optimized to minimize the reconstruction error. Additionally, a regularization term, the Kullback-Leibler divergence, is introduced to ensure that the learned latent space distribution closely aligns with a standard Gaussian distribution. This regularization term contributes to the overall objective of refining the latent space representation in the VAE framework.

VAE is pivotal for HDT in IoT-healthcare. The first key function is compressing and generating complex health data. Additionally, it is also instrumental in anomaly detection, aiding in the early identification of health issues. Moreover, VAE can support health prediction by simulating disease progression \cite{430}. For instance, a VAE-based approach was introduced to reconstruct complete body movements using input signals from the PT's head-mounted device (HMD) \cite{325}. Specifically, the encoder took the input signals and mapped it into a latent space. Then, this latent space representation was  used by the decoder to reconstruct the complete body movements. This method enabled the digital reconstruction of full-body motion based on signals from the head and hands. Similar VAE-based human motion generation was also presented in \cite{433}. These proved beneficial for accurately modeling human motions in scenarios where motion data obtained from the PT may inadequately represent complete body movements, thereby aiding motion monitoring endeavors. Furthermore, a VAE-based method was proposed to predict potential subsequent steps in the clinical measurement trajectory of a patient encountering an ischemic stroke \cite{249}. The model was trained on data from 1216 ischemic stroke patients in the medical information mart for intensive care-IV database. This approach holds promise in simulating disease progression within HDT, thereby assisting in customizing treatment plans tailored to individual patients.

\emph{Transformer}:
Transformer excels at capturing contextual information and long-distance dependencies in text~\textcolor{blue}{\cite{034}}. The architecture used in this model is an encoder-decoder structure, as illustrated in Fig. \ref{Gen}. The encoder employs a bidirectional information propagation process to comprehend the input text. The decoder, found in most transformer architectures, generates words sequentially. This type of decoder is commonly referred to as an autoregressive decoder.
With this architecture, embedding with self-attention mechanisms, it can effectively process the relevant information in the input sequence, making the generated text more accurate, coherent, and able to consider more contextual information.

Transformer is well-matched for HDT in IoT-healthcare by adeptly managing sequential health data, such as EHRs and sensor information. The proficiency in sequence-to-sequence tasks allows it to accurately interpret input sequences and generate corresponding outputs. Furthermore, thanks to the multi-head attention mechanism, transformers can masterfully navigate the intricate web of relationships among diverse health data types, e.g., clinical notes and lab results, enhancing their utility in complex healthcare environments \cite{431}. For example, the state-of-the-art language generation model, generative pre-trained transformer (GPT), has been incorporated into EHR workflows to autonomously respond to patients' healthcare inquiries \cite{233}. This model can be visualized as an HDT representing a doctor, equipped with extensive medical knowledge, capable of addressing patient queries. Additionally, a pre-trained transformer-based approach has been proposed to concurrently learn gene and cell embeddings from a vast amount of single-cell sequencing data. The model further leveraged attention mechanism to capture intricate gene-to-gene interactions at the single-cell level \cite{278}. This approach holds promise in digitally modeling diverse facets of cellular processes within HDT, offering insights into personalized responses to treatments.

\begin{figure*}[!t]
	\centering
	\includegraphics[width=1.01\textwidth]{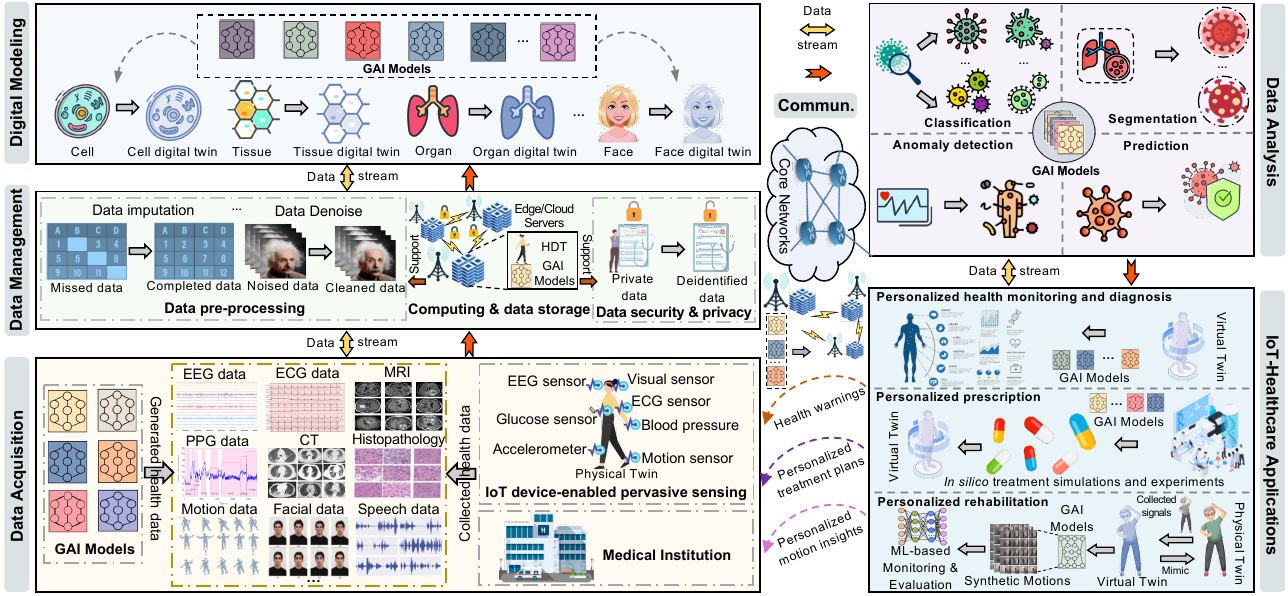} \\
	\caption{The framework of GAI-driven HDT. It includes the GAI-enabled data acquisition, data management, data modeling and data analysis. With the implementation of them, the GAI-driven HDT can be applied in IoT-healthcare, including personalized health monitoring and diagnosis, personalized prescription, and personalized rehabilitation.}\label{fram}
\end{figure*}

\emph{Diffusion model:}
Unlike GANs and VAEs, the diffusion model employs a series of sequential transformations on the input distribution~\cite{du2023diffusion,du2023beyond}. Specifically, as depicted in Fig. \ref{Gen}, this model constructs a Markov chain comprising diffusion steps where noise is incrementally introduced to the input data. Subsequently, a reverse process is implemented, gradually removing noise from the distribution to generate the desired data samples. This inverse method transforms noise distribution back to the original data distribution through a gradual denoising process.

Diffusion model is particularly applicable to HDT in IoT-healthcare. It can recreate complex, high-dimensional distributions of data learned from extensive relevant health datasets. This ability enables the diffusion model to capture the intricacies and variations of human health states and processes, thereby generating highly accurate and individual-specific data involved in HDT \cite{83}. For instance, a diffusion model-based approach was proposed to generate individual electroencephalogram (EEG) data \cite{314}. Specifically, the model was trained on maps of electrode-frequency distribution, which were created from extensive EEG datasets labeled with emotions. It can help solve the EEG data scarcity issue in HDT, and enable HDT to digitally model brain activity patterns, supporting the neurological health monitoring. Additionally, by collaborating with an assistive modality embedding as prior information to diffusion model formulation, a diffusion model-based approach was proposed for positron emission tomography (PET) denoising \cite{296}. It can be used to denoise the acquired data with noise, providing cleaned data to HDT for better personalized healthcare monitoring and diagnostic accuracy. Besides, a diffusion-based unconditional 3D human generative model was proposed to 3D human generation \cite{432}. To be more specific, it learned from 2D human images, leveraging the superior modeling and editing capabilities of structured 2D latents over 1D. The method combined a structured 3D-aware auto-decoder with a latent diffusion model to manage the structured 2D latent space effectively. This approach can facilitate the precise digital modeling of 3D human appearances within HDT.

In addition to the aforementioned models, other GAI models, such as normalizing flows and score-based generative models, have also been effectively implemented in HDT for IoT-healthcare \cite{294, 321}.

\subsection{Framework of GAI-driven HDT}

In this subsection, we give an overview of the framework of GAI-driven HDT, as shown in Fig. \ref{fram}.

Data acquisition component is significantly crucial for HDT, which is the fuel of HDT.
The required substantial health data, such as EEG, ECG, and magnetic resonance imaging (MRI) data, for HDT commonly collected from the IoT-enabled pervasive sensing and medical institutions \cite{du2023semantic}. However, these traditional data acquisitions methods are usually inefficient due to various factors \cite{412, 413}. To this end, GAI can generate ultra-realistic health data based on the collected data for enriching the datasets.

The communication component plays a bridge role in HDT, which is responsible for bi-directional data transmissions between the physical and digital worlds, such as the transmission of the collected data and the feedback in the informal world. However, these data are usually large-scale, multi-modal, and time-sensitive, which are hard to fully met by traditional communication systems \cite{408}. To this end, GAI-enabled communication, such as GAI-enabled semantic communication \cite{246, 385, 386} and cross-modal communication \cite{332}, can be applied to support communication in HDT by generating the transmitted data at the received sides, enhancing the data transmission performances.

Data management component is the core of HDT, where each component will interact with it for data access. Data from both physical and digital worlds, including collected, generated, and simulated ones, are large-scale, leakage-sensitive, and complex. To well manage these data, effective pre-processing, as well as security and privacy schemes are needed. To this end, GAI can be used in data imputation, data denoise, etc., for data pre-processing. Besides, GAI can generate deidentified data while preserving all the patterns from the original data for protecting private data.

Digital modeling component is responsible for the human digitalization procedure in HDT. The digitalization procedure models the high-fidelity human body based on the collected data. However, traditional digital modeling methods highly rely on accurate simulation parameters, which is hard to achieve. To this end, GAI, with its robust generation capabilities, can be adapted to model humans digitally, including cells, tissues, organs, etc.

Data analysis component is essential in HDT, which is responsible for analyzing data in HDT for driving the IoT-healthcare applications. GAI with its strong analysis capabilities can be used in data classification for classifying diseases, data segmentation for obtaining key disease information, anomaly detection for identifying abnormal status, and prediction for predicting health status.

Based on the built GAI-driven HDT described above, it will significantly enhance the IoT-healthcare. It can be used in personalized health monitoring and diagnosis, prescription, and rehabilitation, acting as the intelligent and vivid human digital testbed.

In the following sections, we will delve into the implementation of GAI-driven HDT in IoT-healthcare.

\section{GAI-driven HDT Implementation}\label{GAIforHDT}

\begin{table*}[htbp]
	\centering
	\caption{Summary of GAI-based Approach for Data Acquisition in HDT.}\label{acqui}
	\includegraphics[width=0.93\textwidth]{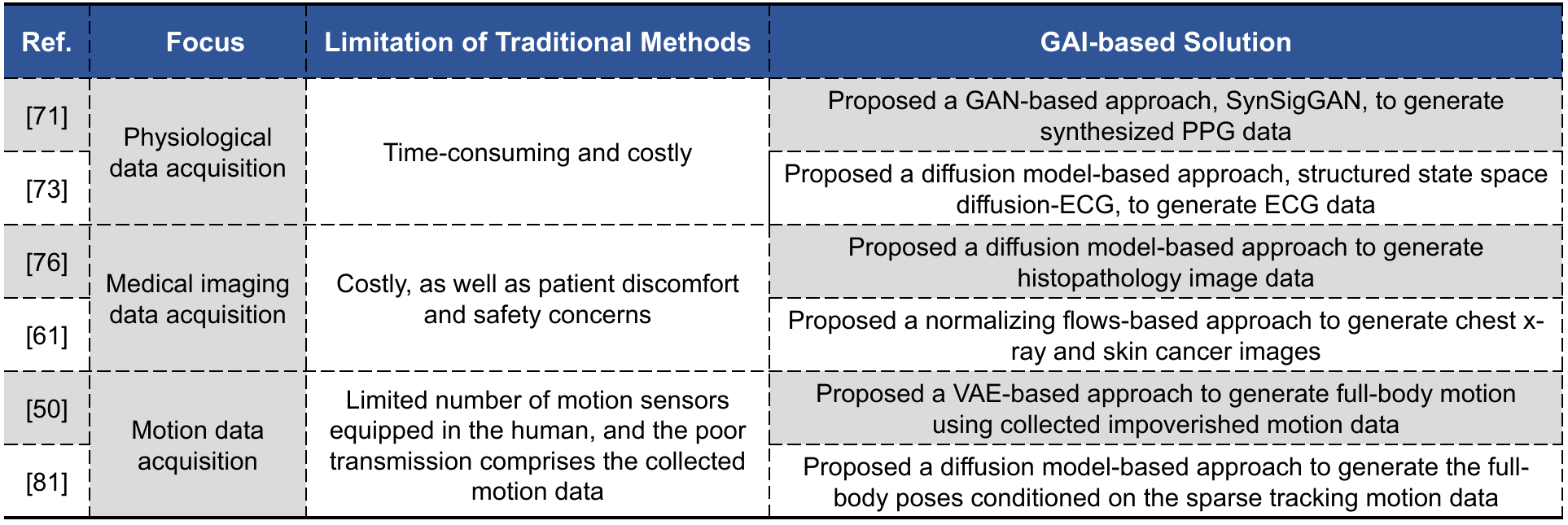}
\end{table*}

\subsection{GAI-enabled Data Acquisition}

\begin{figure*}[!t]
	\centering
	\includegraphics[width=1\textwidth]{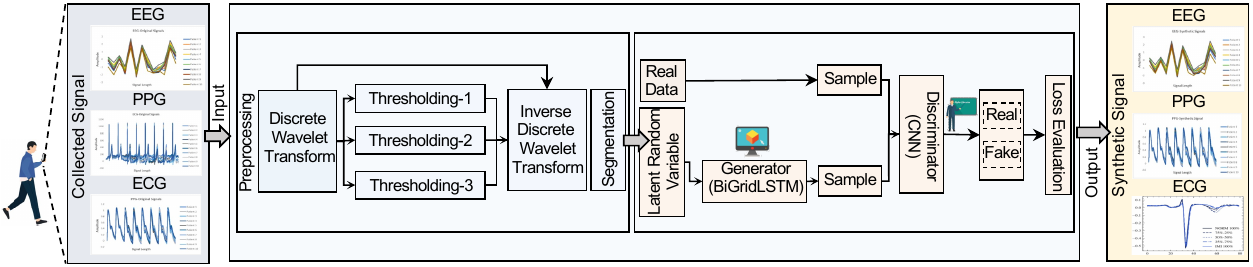} \\
	\caption{Overview of the GAN-based biomedical signal synthesis, SynsigGAN, proposed in [71]. The collected signals proceed through the preprocessing stage, eliminating noise and refining the signals using discrete wavelet transform, thresholding, and inverse discrete wavelet transform. After preprocessing, the signals are forwarded to the segmentation stage that uses the Z-score to solve the amplitude scaling problem and eliminate offset. Next is the GAN, which takes in the segmented signals and generates synthetic biomedical signals using bidirectional grid long short-term memory for generator network and convolutional neural network for the discriminator. Finally, SynsigGAN outputs the synthesized biomedical signals.}\label{synsig}
\end{figure*}

Data forms the cornerstone of HDT development and service provision \cite{4}. Commonly employed data acquisition methods for HDT primarily stem from diverse sources, including medical institutions \cite{39}, as well as both non-invasive and invasive sensors equipped in PTs \cite{27}. Nevertheless, these methods present challenges due to their time-consuming, costly, intrusive nature, and limited scalability. This restricts acquiring extensive individual-level data essential for robust HDT development and comprehensive service delivery. GAI can significantly assist the data acquisition in HDT, by offering diverse and highly realistic synthetic data. In the following, we introduce several common synthetic HDT data generated by GAI, including synthetic physiological, medical imaging and motion data, as summarized in Table \ref{acqui}.

The physiological data, such as ECG, EEG, and photoplethysmogram (PPG), hold significant importance for HDT. 
By generating synthetic physiological data, GAI mitigates the limitations of traditional physiological data acquisition methods, reducing costs and time associated with collecting extensive real-world physiological data. For instance, 
SynSigGAN proposed by Hazra et al. in \cite{316} can generate synthesized PPG data, as shown in Fig. \ref{synsig}. In SynSigGAN, the bidirectional grid long short-term memory (LSTM) and the convolutional neural network (CNN) had been used for generator network and discriminator network, respectively, and it was trained on  BIDMC PPG and respiration datasets \cite{367}. In addition, diffusion model-based approaches also applied in physiological data generation. 
Alcaraz et al. in \cite{288} designed a diffusion model-based approach, called structured state space diffusion-ECG (SSSD-ECG), for generating synthetic 12-lead ECG data. SSSD-ECG was built on the SSSD$^{S4}$ model architecture \cite{318} and trained on PTB-XL dataset \cite{317}, which was a publicly available collection of clinical 12-lead ECG data comprising 21,837 records from 18,885 patients. These highly realistic synthetic physiological data can significantly amplify the physiological datasets that required by HDT, to enhance the performance.

\begin{figure*}[!t]
	\centering
	\includegraphics[width=1\textwidth]{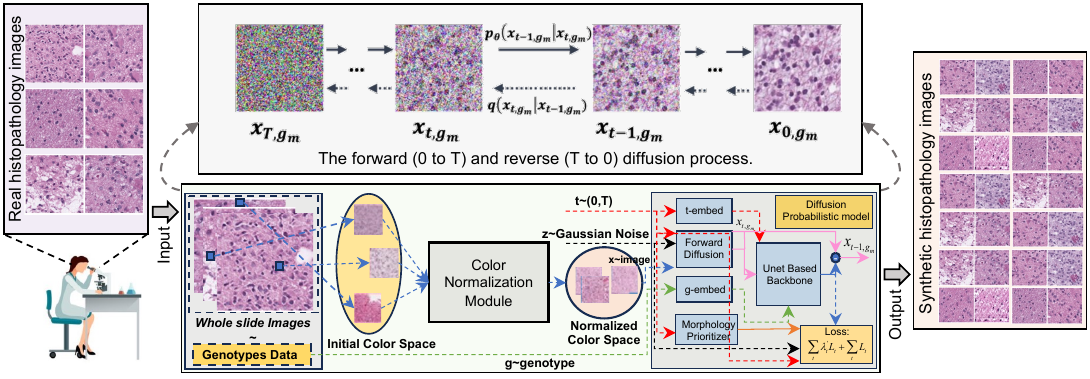} \\
	\caption{Overview of the diffusion model-based histopathology image synthesis approach proposed in [76].  The real histopathology images are extracted the genotype information firstly, then it is used as a conditional input to the diffusion probabilistic model, which generates synthetic histopathology images that are tailored to specific genotypes.}\label{heal}
\end{figure*}

The medical imaging data hold significant importance for HDT. Medical imaging provides detailed insights into a PT's anatomy, allowing for a precise digital representation. This information aids in creating accurate simulations of physiological structures and functions within the VT. Besides, by integrating this data, the VT can simulate and predict the progression of diseases. However, existing medical imaging acquisition methods rely on medical imaging devices, such as MRI or computed tomography (CT) scanners, which encounter high costs, patient discomfort, and safety concerns. To this end, GAI can generate synthetic medical images, augmenting existing datasets without additional patient scans. For instance, Moghadam et al. in \cite{319} proposed a diffusion model-based approach for the synthesis of histopathology images, as shown in Fig. \ref{heal}. Specifically, the authors used color normalization to force the diffusion model-based approach to learn morphological patterns, and used perception prioritized weighting, aiming to prioritize focusing on diffusion stages with more important structural histopathology contents. Experimental results showed that the proposed approach outperformed the GAN-based approach proposed in \cite{368} by generating high quality histopathology images of brain cancer. While diffusion model and GAN made remarkable progress in medical image generation, they cannot explicitly learn the probability density function of the input data and are highly sensitive to the hyperparameter selections. To mitigate these issues, Hajij et al. in \cite{321} investigated normalizing flows (NFs) based approach as an alternative for synthesizing medical images. Particularly, the authors trained a RealNVP \cite{322}, a popular NF model for medical image synthesis, on two medical imaging datasets: chest X-ray \cite{323} and skin cancer \cite{324}. The experimental results showed that the NF-based medical image synthesis approach is an attractive alternative to GAN-based and diffusion model-based approaches.

\begin{figure*}[!t]
	\centering
	\includegraphics[width=1\textwidth]{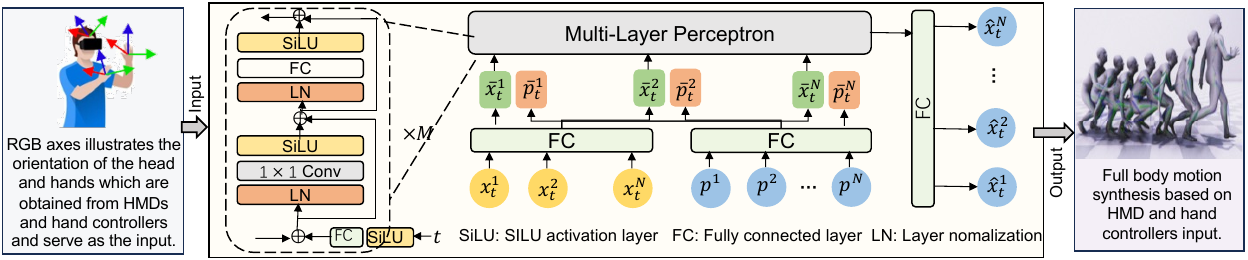} \\
	\caption{Overview of AGRoL proposed in [81]. It takes the orientations of the head and hands from HMDs and hand controllers as the input. These input processed by AGRoL. The architecture of AGRoL is presented in the middle of this figure, where $t$ is the noising step. $x_{t}^{1:N}$ denotes the motion sequence of length $N$ at step $t$, which is pure Gaussian noises when $t=0$. $p^{1:N}$ denotes the sparse upper body signals of length $N$. $\hat{x}_{t}^{1:N}$ denotes
		the denoised motion sequence at step $t$. The output is the synthesized full body motion.}\label{agrol}
\end{figure*}

The motion data is significant for HDT, which can be used to model and simulate a PT's entire body posture and movements in the digital space. Motion modeling in HDT can enable critical HDT services, such as motion monitoring during rehabilitation and injury prediction during exercise. Existing motion data acquisition mainly through wearable devices \cite{4}, which has several drawbacks hindering the complete and accurate motion data acquisition. For example,  humans may not always be equipped with a large number of motion sensors for acquiring comprehensive motion data, which results in the acquired data can only characterize partial motion. Moreover, issues related to signal transmission, such as poor connectivity, signal interference, or obstructions caused by complex human mobility, can result in data loss or corruption during transmission. In this regard, GAI has been successfully applied in motion data synthesis, which can be a promising solution. For instance, Dittadi et al. in \cite{325} proposed a VAE-based approach to generate full-body motion based on an impoverished control signal coming from HMDs. Specifically, to reconstruct the articulated poses of a human skeleton from noisy streams of head and hand pose, the authors proposed VAE-based approach decomposed the problem into a generative model of human pose, with an inference model that mapped input signals into the learned latent embedding. Experiment results showed that the proposed approach can faithfully reconstruct the walking motion of the person wearing an HMD. Additionally, to accelerate the motion generation rate to meet online application requirements, Du et al. in \cite{326} proposed a diffusion model-based approach, called avatars grow legs (AGRoL), to generate the full-body poses conditioned on the sparse tracking signals from HMDs, as shown in Fig. \ref{agrol}. To enable gradual denoising and produce smooth motion sequences, the authors proposed a block-wise injection scheme that added diffusion timestep embedding before every intermediate neural network block. With this timestep embedding strategy, AGRoL achieved SOTA performance on the full-body motion synthesis task without any extra losses that were commonly used in other motion prediction methods. In addition, due to the lightweight architecture, AGRoL can generate realistic, smooth motions while achieving real-time inference speed, making it suitable for online applications.

In summary, GAI offers promising solutions to address the limitations of current data acquisition methods in HDT by generating diverse and high-realistic datasets.

\subsection{GAI-enabled Communication}

HDT relies on real-time data transmission to keep synchronization between any PT-VT pair to ensure high-fidelity of VT \cite{394, 422, 423}. This synchronization is, however, data-driven and delay-sensitive. Furthermore, data acquired in the physical world is often massive and complex. In addition to real-time synchronization, interacting with VTs involves more complex information that needs to be transmitted between the PT and users in the physical world. Multi-modal information, such as 3D virtual items, text, images, haptic signals, among others, needs to be transmitted in HDT under various applications, such as the virtual surgery, to enhance the immersive experience. These specific characteristics place a significant burden on current communication networks \cite{402}. To address these issues, this subsection delves into the GAI-aid semantic communication \cite{385} and cross-modal communication \cite{332} for communication in HDT, as summarized in Table \ref{comm}.

\begin{table*}[htbp]
	\centering
	\caption{Summary of GAI-based Approach for Communication in HDT.}\label{comm}
	\includegraphics[width=0.94\textwidth]{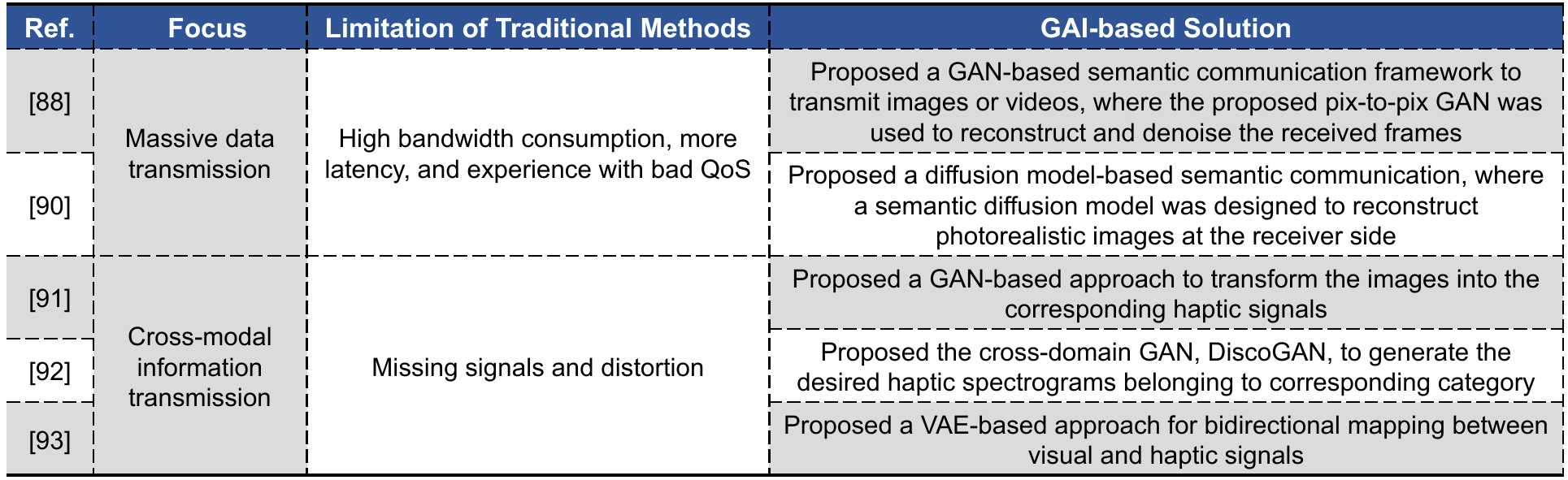}
\end{table*}

Semantic communication is expected to enable the data transmission between the PT and VT pair in HDT, tackling the challenges of unnecessary transmission of vast amounts that cause high bandwidth consumption, more latency, and experience with bad quality of service (QoS) by only transmitting meaningful and task-oriented information extracted from the original information~\cite{yang2022semantic,du2023user}. Generally, semantic communication extracts the ``meaning'' of any transmitted information at the transmitter and encodes the extracted features. Then, this semantic information is transmitted to the intending receiver and is ``interrupted'' and decoded by the receiver. GAI with its creativity is applied in receiver side to reconstruct the original information from the received semantic information. For instance, Raha et al. in \cite{328} proposed a GAN-aid semantic communication framework for transmitting images or videos. Specifically, by considering resource limitations inherent in edge devices, such as HMDs, and the need for low-latency transmission in HDT applications, such as the real-time PT-VT synchronization task, the authors utilized a lightweight mobile segment anything model \cite{329} for essential semantic information extraction from the images or videos. Then, on the receiver side, the authors proposed a pix-to-pix GAN approach to reconstruct and denoise the received semantic frames. The simulation results showed that the proposed framework can reduce up to 93.45\% of the communication cost while maintaining the original information. In addition to GAN-aid semantic communication, the diffusion model with strong synthesizing multimedia content abilities has also been applied in this field. For example, Grassucci et al. in \cite{331} introduced a semantic diffusion model designed to reconstruct photorealistic images at the receiver side. This model was trained using noisy semantics and incorporated a fast denoising semantic block to enhance the quality of inferred images. Consequently, the receiver can reconstruct semantically-consistent samples from the compressed semantic images transmitted by the sender over a noisy channel.
These methods can be applied in HDT for rehabilitation training to reduce bandwidth usage and latency. The cameras are utilized to capture the patient's motion, and from the captured extensive images or videos, semantic features are extracted. Subsequently, the highly-compressed semantic information is transmitted to the digital world, where GAI is employed to regenerate the images or videos, and then, synchronizing the motion of the corresponding VT. This approach enables real-time monitoring and analysis of rehabilitation conditions \cite{8,du2023user}.

The interactions in HDT applications, such as virtual surgery, commonly involve with audio, visual and haptic signals to provide users with human interactive and immersive experiences. To support such interactions within HDT, cross-modal communication \cite{332}, which involves collaborative audio-visual and haptic interactions, presents a promising solution. It adeptly resolves the distinct requirements among these modalities when they coexist. However, long transmission distance between the physical and digital world or possible poor network conditions may result in missing signals and distortion, negatively impacting the users' experiences. To this end, researchers have explored using GAI to enable cross-modal signal reconstruction within the cross-modal communication to compensate for defective signals and achieve high-fidelity communication. For instance, by fully using of the correlation between image and haptic modalities, Liu et al. in \cite{334} proposed a GAN-based approach to transform the images into the corresponding haptic signals. Specifically, the authors extract the image features firstly to obtain the required category information. Then, they adopted the cross-domain GAN, DiscoGAN \cite{335}, to generate the desired haptic spectrograms belonging to that category. In addition to cross-modal unidirectional mapping illustrated above, the cross-modal bidirectional mapping is investigated to enhance the cross-modal signal reconstruction. For example, Fang et al. in \cite{336} proposed a VAE-based approach for bidirectional mapping between visual and haptic signals. Specifically, the authors firstly adopted the visual VAE model and the haptic VAE for compressing visual and haptic data, respectively. Then, they employed a conditional flow model to connect the latent feature spaces of these two VAE models. The forward process of the flow model was the mapping from the haptic to the visual latent feature space, while the reverse process was the mapping from the visual to the haptic latent feature space. Based on this, the cross-modal bidirectional mapping between visual and haptic signals can be successfully implemented on one model. These GAI-aid cross-modal communication approaches are crucial for cross-modal interactions in HDT. For example, during virtual acupuncture training, when a doctor interacts with a patient's VT, haptic signals may be missed or distorted due to the degraded channel conditions, while the visual signals are successfully received by the doctor's VR device \cite{337}. To provide the immersive experience, the GAI-aid haptic signal reconstruction approach can be utilized to reconstruct the haptic signals from the visual signals.

\subsection{GAI-enabled Data Management}
Data management is a crucial step in the successful implementation of HDT. First, the collected multi-source data in HDT may have the characteristics of heterogeneity, multi-scale and high noises. Hence, as the critical step in data management, data pre-processing is indispensable in HDT, such that issues like missing data and noise data can be properly handled, and thereby providing high-quality data for downstream tasks in HDT \cite{4}. Moreover, the data in HDT are highly sensitive, especially for individual-level data, and any leakage may result in serious ethical and moral concerns. Therefore, effective security and privacy schemes are imperative to protect data in HDT. In the following, we will discuss the application of GAI in data pre-processing, followed by its application in security and privacy schemes, as summarized in Table \ref{datama}.

\begin{table*}[htbp]
	\centering
	\caption{Summary of GAI-based Approach for Data Management in HDT.}\label{datama}
	\includegraphics[width=0.9\textwidth]{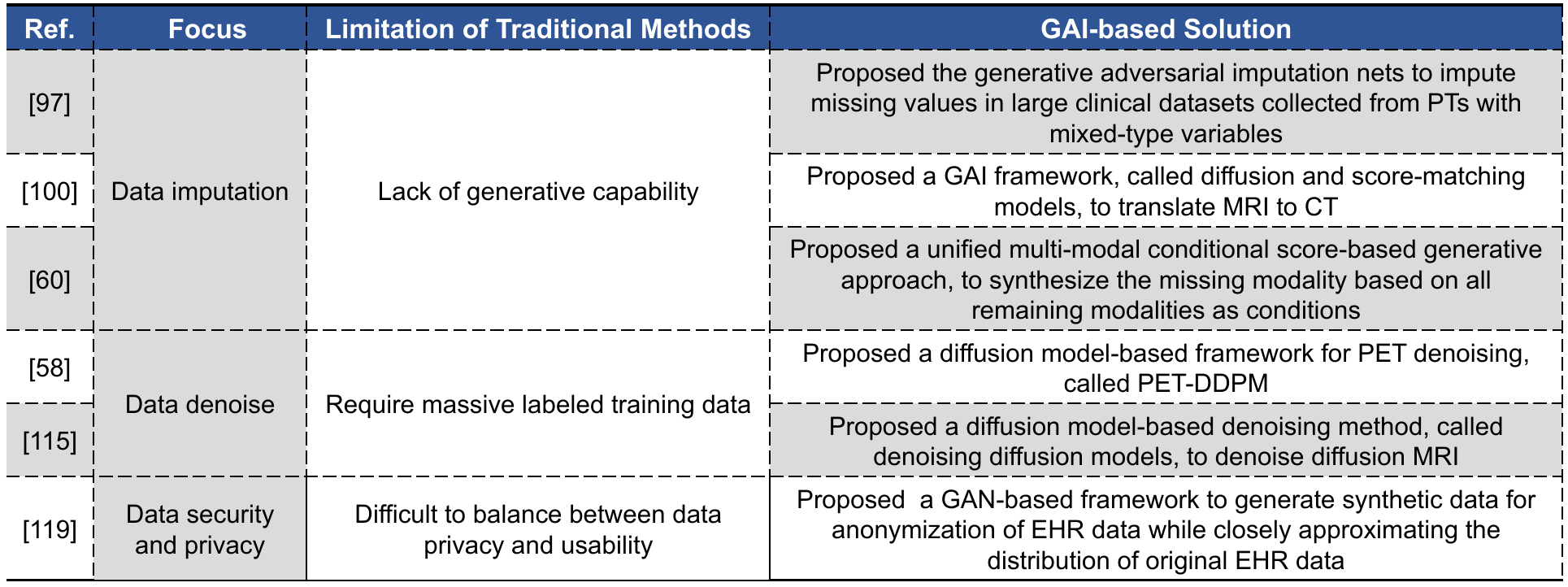}
\end{table*}

Traditional data imputation methods, such as K-nearest neighbors-based \cite{40} and deep learning-based imputation \cite{358}, often rely on existing observed data for modeling and prediction, and cannot directly generate new data samples. This will make it hard for traditional data imputation methods to handle missing severe data issues, such as data modality missing. To this end, GAI, with its powerful generative capabilities, can effectively handle the missing data issues in HDT. For instance, with the potential of substituting missing data accurately and efficiently, Dong et al. in \cite{84} used GAN to impute missing values in large clinical datasets collected from PTs with mixed-type variables. Specifically, the method adopted by authors was generative adversarial imputation nets (GAIN), where the generator observed some components of a real clinical data vector, imputed the missing components conditioned on what was actually observed, and outputs a complete clinical data vector. The discriminator then took a completed clinical data vector and attempted to determine which components were observed and which were imputed. Additionally, to ensure that discriminator forced generator to learn the desired data distribution, the authors provided discriminator with some additional information in the form of a hint vector. The accuracy of GAIN was measured by imputation error, defined as the difference between the imputed values and real value. Specifically, normalized root mean square error (NRMSE) was used to assess difference for continuous variables, while proportion of falsely classified (PFC) was used for categorical variables. The experimental results showed that GAIN outperformed the traditional imputation models, MICE \cite{365} and missForest \cite{366}, in terms of NRMSE and PFC in the imputation of missing data in clinical datasets, under the missingness rate of 20\% and 50\%. Acquiring multi-modality data is crucial for the implementation of HDT. However, the missing modalities may happen in some conditions, such as failure in data transmission. Fortunately, diffusion models have shown favorable results for generating missing modalities utilizing cross-modalities and producing ones using other modality types. For instance, Lyu et al. in \cite{291} proposed a GAI framework, called diffusion and score-matching models, which took advantage of the recently introduced denoising diffusion probabilistic models (DDPMs) \cite{292} and score-based diffusion models \cite{293}, for translating MRI to CT. Specifically, they presented conditional DDPM and conditional stochastic differential equation (SDE) \cite{369}, where their reverse process was conditioned on T2w MRI images. The authors adopted the DDPM and SDE with three different sampling methods, namely Euler-Maruyama, Prediction-Corrector, and probability flow ordinary differential equation. In their experiments, they utilized the structural structural similarly index measure (SSIM) and peak signal-to-noise ratio (PSNR) metrics to evaluate their quality of translating MRI to CT. The datasets used in this experiment is Gold Atlas male pelvis dataset \cite{370}, and the results showed that the proposed method achieved SSIM = 0.86 and PSNR = 27.31, which was superior than both GAN (SSIM = 0.83, PSNR =  26.84) and CNN-based methods (SSIM = 0.84, PSNR = 26.98) \cite{371}.
Similarly, to cope with the missing modality issue, Meng et al. in \cite{294} proposed a unified multi-modal conditional score-based generative approach (UMM-CSGM), which synthesized the missing modality based on all remaining modalities as conditions. The proposed model was a conditional SDE format, employing only a score-based network to learn different cross-modal conditional distributions. The experimental results showed that the UMM-CSGM could generate missing-modality images with higher fidelity and structural information of the brain tissue compared to GAN-based methods \cite{372, 373, 374, 375, 376}.

Noise data is a common issue during the data acquisition in HDT. Noise reduces the data quality and is especially significant when the point of interests are minor and have relatively low contrast, which hinders the downstream tasks in HDT \cite{295}. Traditional data denoise methods, such as CNN-based \cite{359} and U-Net-based  methods \cite{360}, highly rely on a significant amount of labeled training data to build the denoising models. However, obtaining such data is challenging, especially at the individual level. To this end, with the strong generative abilities of GAI, it is convenient for diverse denoising problems in HDT data. For instance, PET is a medical imaging technique used to detect metabolic activities within the human body. The information provided by PET can be used to update or calibrate the VT models. However, due to various physical degradation factors, PET often suffers from low SNR and resolution. To denoise the PET images, a DDPM-based method has been proposed for PET image denoising. Gong et al. in \cite{296} proposed the DDPM-based framework for PET denoising, called PET-DDPM, which collaborated with an assistive modality embedding as prior information to DDPM formulation. Quantitative results demonstrated that PET-DDPM with SSIM = 0.91 and PSNR = 33. 84 outperformed U-Net based denoising networks with SSIM = 0.87 and PSNR = 33.32 \cite{377}, showcasing its superior performance in PET denoising. For denoising MRI that with severely SNR-limited, Xiang et al. in \cite{297} proposed a self-supervised denoising method, called denoising diffusion models for denoising diffusion MRI (DDM$^{2}$). Specifically, their approach consisted of three stages, which integrated statistic-based denoising theory into diffusion models and performed denoising through conditional generation. During inference, they represented input noisy measurements as a sample from an intermediate posterior distribution within the diffusion Markov chain. The experiment results showed that DDM$^{2}$ demonstrated superior denoising performances ascertained with clinically-relevant visual qualitative and quantitative metrics than Noise2Score (N2S) \cite{434} and other denoising methods. Specifically, DDM$^{2}$ not only suppressed noise the most but also restored anatomical details commented by two neuroradiologists. Additionally, compared to N2S, DDM$^{2}$ demonstrated an average of 0.95 and 0.93 improvement on SNR and Contrast-to-Noise Ratio, respectively.

\begin{figure}[!t]
	\centering
	\includegraphics[width=0.99\columnwidth]{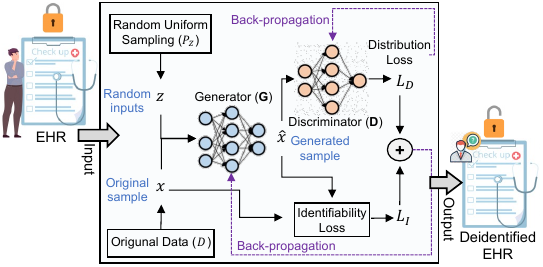} \\
	\caption{Block diagram of ADS-GAN for individual EHR anonymization [119]. The generator uses original sample ($x$) and random vector ($z$) to generate sample ($\hat{x}$). The summation of distribution loss (LD) and identifiability loss (LI) is back-propagated to the generator. Both the generator and discriminator are implemented with multi-layer perceptrons.}\label{ADS}
\end{figure}

HDT environments may suffer from various security threats, such as eavesdropping attacks in data transmission between PTs and VTs causing privacy leakage. A typical privacy protection mechanism for HDT data is anonymization \cite{361, 362}. However, this method can easily compromise the data distribution, and it is hard to balance data privacy and usability. To this end, GAI can generate synthesized HDT data, which maximally preserves original data distribution while guaranteeing data security and privacy. For instance, to lower the risk of breaching individual confidentiality during data sharing in HDT, Yoon et al. in \cite{298} proposed a GAN-based framework, called anonymization through data synthesis using GAN (ADS-GAN), for generating synthetic data that closely approximates the distribution of variables in an original EHR dataset, achieving anonymization of EHR, as shown in Fig. \ref{ADS}. Specifically, ADS-GAN generated the synthetic data conditioned on the original data, and different from traditional conditional GAN framework, conditioning variables in ADS-GAN were not pre-determined but instead were optimized from real EHR. Additionally, authors proposed a mathematical definition for identifiability to quantify the degree of anonymization. Particularly, $\epsilon$-identifiability they defined meant that the probability of patients can not identifiable by the synthetic patient data was larger than $\geq 1- \epsilon$, which was attributed to the synthetic patient data were ``different enough" from the real ones. To prevent re-identification by generated synthetic data, authors integrated the constraint, $\epsilon$-identifiability, into ADS-GAN. To this end, they designed a identifiability loss for training discriminator in ADS-GAN. The experiment results showed that ADS-GAN performed superior than traditional conditional GAN frameworks, i.e., PATE-GAN \cite{380} and WGAN-GP \cite{371}. ADS-GAN can be 27\% and 67\% better than PATE-GAN in terms of both Jensen-Shannon Divergence (JSD) and the Wasserstein distance with the same identifiability. While with the same quality of the generated samples, ADS-GAN reduced the identifiability by 41\% compared to WGAN-GP.

In summary, GAI is a promising technique for data management in HDT, handling data pre-processing and security and privacy issues.

\subsection{GAI-enabled Digital Modeling}
In HDT, digital modeling refers to digitally model the PT and virtualized in the digital world based on the acquired data and various digital modeling technologies, forming the VT.

The classic HDT digital modeling technology, mechanistic modeling, has pioneered the integration of biology and physiology domain knowledge to allow robust and accurate modeling \cite{269}. For instance, a human heart DT model has been developed, which consisting of approximately 100 million virtual heart cell patches, with each patch modeled by around 50 equations \cite{65}. The human heart DT accurately represented the interconnected cardiac muscle cells, effectively simulating the transmission of electrical currents through these cells and the subsequent initiation of the heartbeat. However, as the human heart DT, mechanistic modeling requires massive and accurate simulation parameters, which are particularly challenging to acquire from biological entities in the human body, i.e., from molecular to organ level, and are typically limited to only a subset of all available biomolecular \cite{269}. GAI techniques can overcome these challenges, through learning the underlying distribution and sequential or temporal relations of data for modeling. In the following, we provide a survey of the application of GAI in HDT digital modeling, including digital modeling of human cells, tissues and organs, as summarized in Table \ref{model}.

\begin{table*}[htbp]
	\centering
	\caption{Summary of GAI-based Approach for Digital Modeling in HDT.}\label{model}
	\includegraphics[width=0.94\textwidth]{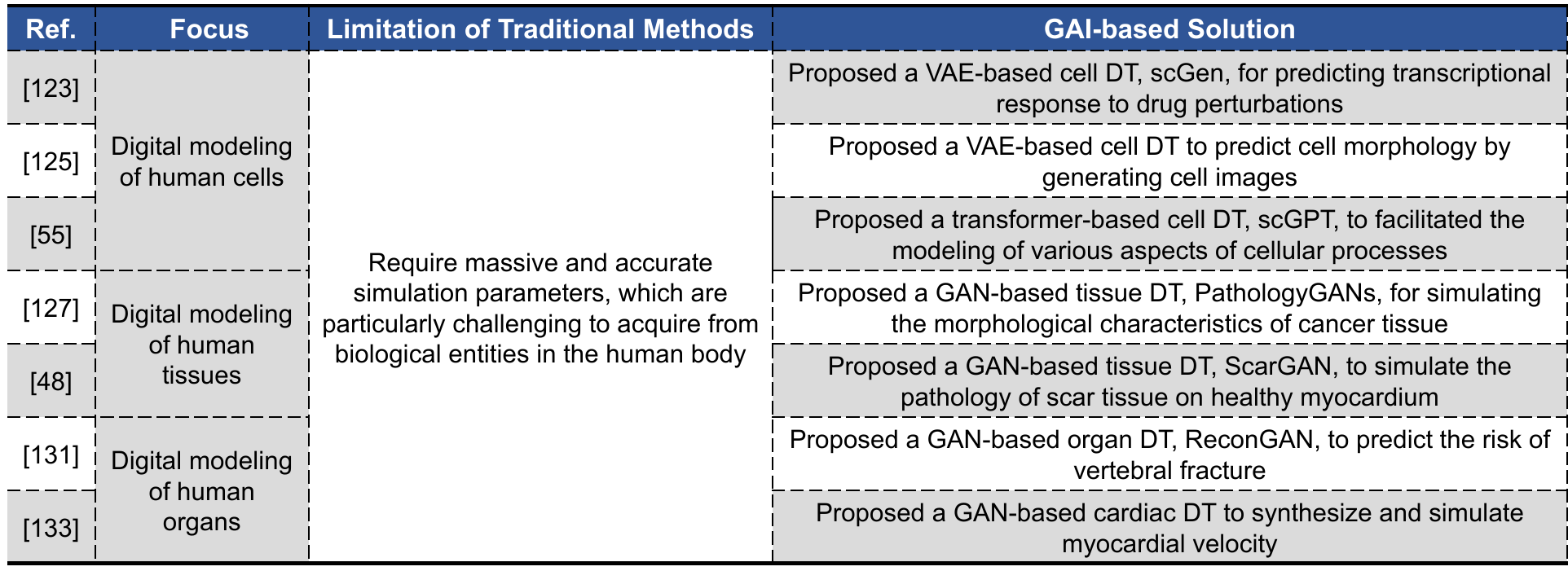}
\end{table*}

\begin{figure*}[!t]
	\centering
	\includegraphics[width=1\textwidth]{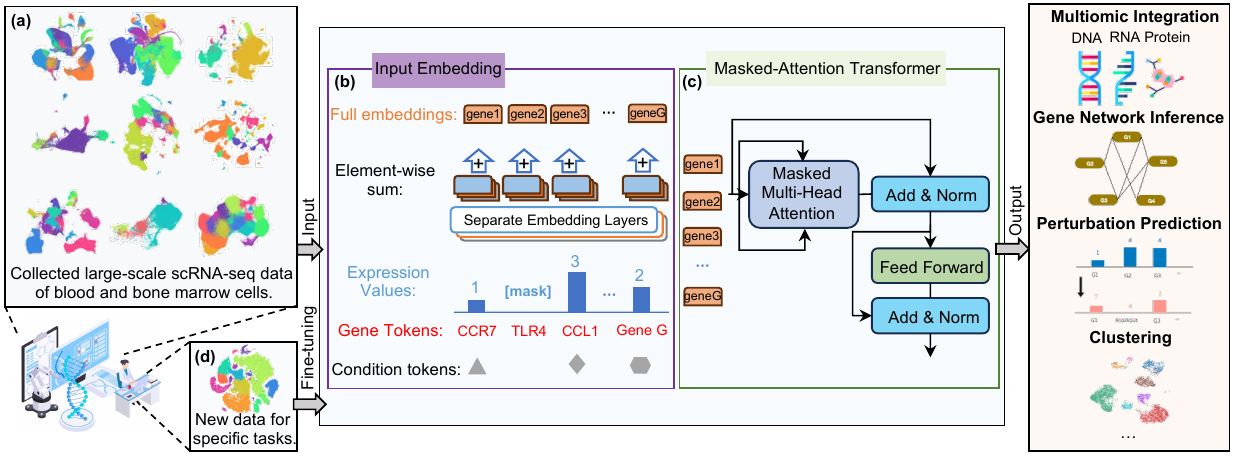} \\
	\caption{Overview of the transformer-based cell DT, scGPT, proposed in [55]. (a) Large-scale scRNA-seq datasets from cell atlas are obtained from medical institutions and input to the input embedding module. (b) The input contains three layers of information, the gene token, expression value, and condition tokens (modality, batch, perturbation condition, et al.). (c) A specially designed mask-attention transformer is proposed to conduct generative pre-training on single-cell sequencing data. (d) For downstream applications (e.g., multiomic integration, gene network inference, perturbation prediction, and clustering), the pre-trained model weights can be finetuned on the new data.}\label{scGPT}
\end{figure*}

First, GAI techniques have shown their strong abilities in digitally modeling human cells. For instance, Lotfollahi et al. in \cite{251} built a VAE-based cell DT, scGen, for predicting transcriptional response to drug perturbations. Specifically, scGen combined VAE and latent space vector arithmetics for high-dimensional sing-cell gene expression data. scGen can accurately model perturbation and infection response of cells across cell types. The results of the simulations showcased that scGen successfully learned cell-type responses, indicating its ability to capture distinguishing features between responding and non-responding genes and cells. Furthermore, the simulation results provided evidence for the enhanced generalization capabilities of scGen compared to mechanistic modeling approaches, which are typically tailored to specific cellular settings. In addition, the authors further developed another VAE-based cell DT, compositional perturbation autoencoder (CPA), to predict cellular response to unseen drugs, drug combinations and dosages in high-throughput screens \cite{252}. CPA combined the interpretability of linear models with the flexibility of deep learning approaches for single‐cell response modeling. The authors envisioned that with the accurate modeling of the cell, CPA will accelerate therapeutic applications using single-cell technologies. Similarly, Donovan-Maiye et al. in \cite{279} developed a VAE-based cell DT to predict cell morphology by generating cell images. Specifically, they employed stacked conditional $\beta$-VAE to first learn a latent representation of cell morphology, and then learn a latent representation of subcellular structure localization which is conditioned on the learned cell morphology under treatment. Inspired by parallels between linguistic constructs and cellular biology, where text comprises words, similarly, cells are defined by genes, pre-trained transformers are envisioned to model the cell DT. Cui et al. in \cite{278} developed a transformer-based cell DT, scGPT, a generative pre-trained foundation model that harnessed the power of pre-trained transformers on a vast amount of single-cell sequencing data, as shown in Fig. \ref{scGPT}. The use of transformers in scGPT enabled the simultaneous learning of gene and cell embeddings, which facilitated the modeling of various aspects of cellular processes. In addition, by leveraging the attention mechanism of transformers, scGPT captures gene-to-gene interactions at the single-cell level, providing an additional layer of interpretability.

\begin{figure}[!t]
	\centering
	\includegraphics[width=0.92\columnwidth]{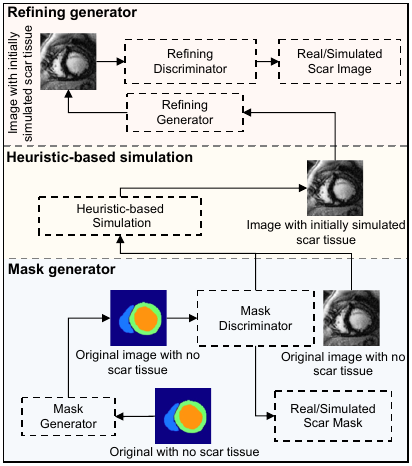} \\
	\caption{Overview of the GAN-based tissue DT, ScarGAN, propoesd in [48]. A mask generator simulates the shape of scar tissuesegmentation mask (left ventricular endo is light blue; left ventricular myo is green; left ventricular endo is orange; scartissue is red); a heuristic-based method provides an initial simulated scar tissue using the simulated shape; a refining generator add details of scar tissue to the image.}\label{ScarGAN}
\end{figure}

GAI has been successfully applied in the digital modeling of human tissues for digital pathology \cite{280}. For instance, Quiros et al. in \cite{281} developed a GAN-based tissue DT, PathologyGANs, for simulating the morphological characteristics of cancer tissue. Specifically, PathologyGANs combined BigGAN \cite{282}, StyleGAN \cite{283}, and relativistic average discriminator \cite{284} to learn representations of entire tissue architecture. Then, they used these characteristics to structure PathologyGANs' latent space (e.g., color, texture, spatial features of cancer and normal cells, and their interaction). Thus, PathologyGANs can generate high-fidelity cancer tissue images from the structured latent space. The simulation showed that the quality of the generated cancer tissue images did not allow pathologists to reliably find differences between real and generated images. It indicated that the proposed GAN-based tissue DT can accurately characterize the features of cancer tissues. 
To reduce the frequent need to collect scans from patients, Lau et al. in \cite{286} built a GAN-based tissue DT, ScarGAN, to simulate the pathology of scar tissue on healthy myocardium, as shown in Fig. \ref{ScarGAN}. Specifically, ScarGAN was based on chained GAN, and the simulation process included 3 steps: i) a mask generator to simulate the shape of the scar tissue; ii) a domain-specific heuristic to produce the initial simulated scar tissue from the mask; iii) a refining generator to add details to the simulated scar tissue. The experimental results conducted by the authors showed that ScarGAN can high realistically simulate scar tissue on normal scans, such that experienced radiologists could not distinguish between real and simulated scar tissue.

GAI has been used in the digital modeling of human organs for computationally reproducing normal and pathological organ function and treatment effect. For instance, Ahmadian et al. in \cite{256} built a GAN-based organ DT, coined ReconGAN, to predict the risk of vertebral fracture (VF). Specifically, ReconGAN consisted of a deep convolutional GAN (DCGAN) \cite{381}, image-processing steps, and finite element (FE) based shape optimization to reconstruct the vertebra model. The synthetic trabecular microstructural models generated by DCGAN were infused into the vertebra cortical shell extracted from the patient's diagnostic CT scans using an FE-based shape optimization approach to achieve a smooth transition between trabecular to cortical regions. The final geometrical model of the vertebra was converted into a high-fidelity FE model to simulate the VF response using a continuum damage model under compression and flexion loading conditions. The experiments implemented by the authors demonstrated that the built GAN-based vertebra DT can accurately simulate and predict the risk of VF in a cancer patient with spinal metastasis. As the core organ of human bodies, the GAI aided human heart digital modeling has also been investigated. For instance, to overcome the long acquisition time and complex acquisition of cardiac data, Xing et al. in \cite{254} developed a GAN-based cardiac DT, hybrid deep learning (HDL), to synthesize and simulate myocardial velocity maps from real-world three-directional CINE multi-slice myocardial velocity mapping (3Dir MVM) data. Specifically, the HDL was featured by a hybrid UNet and a GAN with a foreground-background generation scheme. The experimental results demonstrated that the synthetic 3Dir MVM data generated from the HDL algorithm can accurately and quantitatively assess the cardiac motion in three orthogonal directions of the left ventricle.

In summary, the data-driven GAI based digital modeling can effectively overcome the shortcomings of mechanistic modeling, including requiring accurate simulation parameters and sufficient prior domain knowledge. Therefore, GAI is a promising technique for digital modeling of HDT.

\subsection{GAI-enabled Data Analysis}
Data analysis is the crucial component in HDT, which analyzes the data collected from the physical world and the data generated in the digital world. The results from data analysis are essential information in providing HDT services, such as diagnosis and prescription. Traditional data analysis methods, such as CNN-based \cite{364} and U-Net based methods \cite{363}, highly rely on a large amount of labeled data and difficult in handling high-dimensional HDT data. To this end, GAI has showcased remarkable potential in data analysis, including classification, segmentation, anomaly detection and prediction, as summarized in Table \ref{analy}.

\begin{table*}[htbp]
	\centering
	\caption{Summary of GAI-based Approach for Data Analysis in HDT.}\label{analy}
	\includegraphics[width=0.9\textwidth]{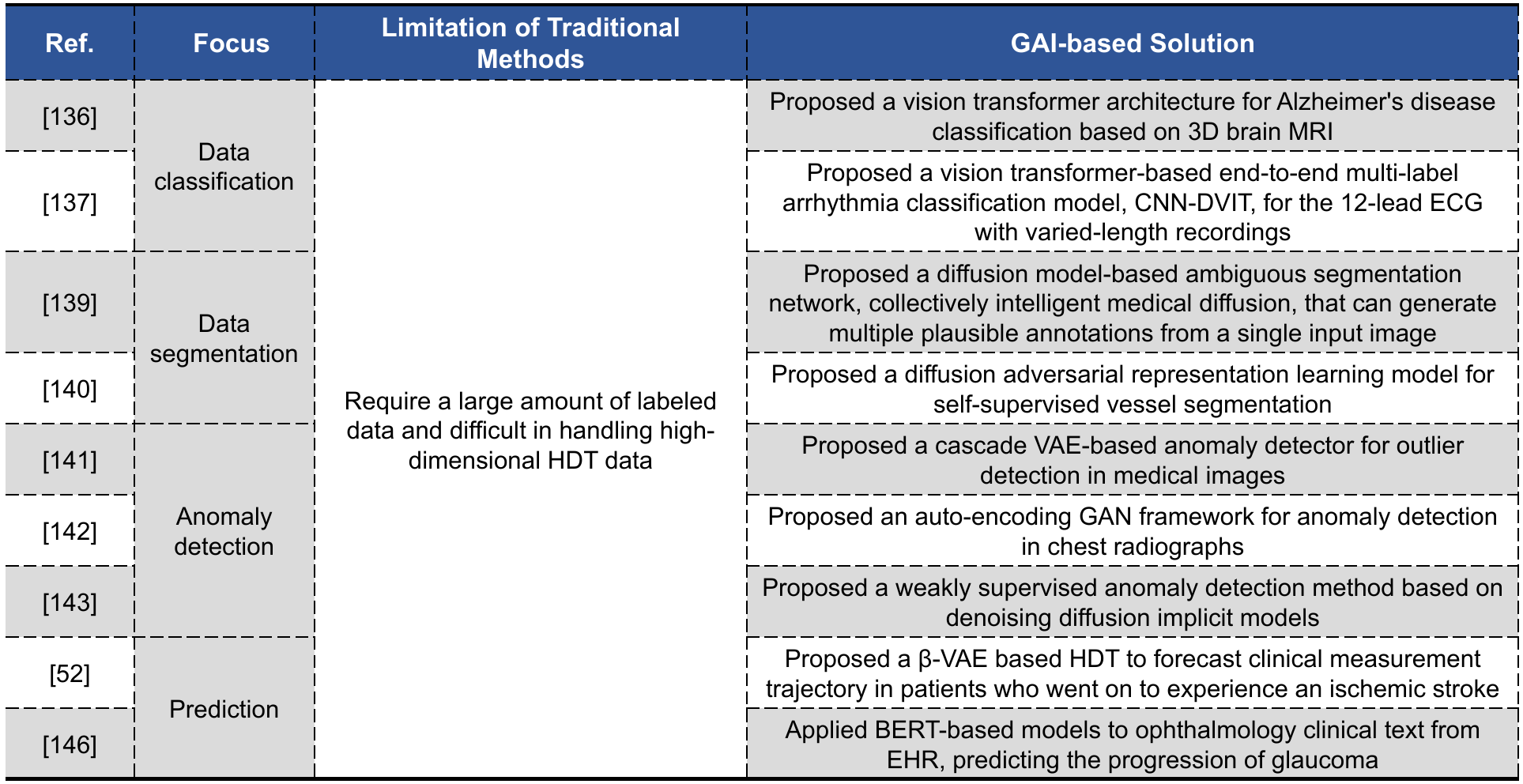}
\end{table*}

\begin{figure}[!t]
	\centering
	\includegraphics[width=0.76\columnwidth]{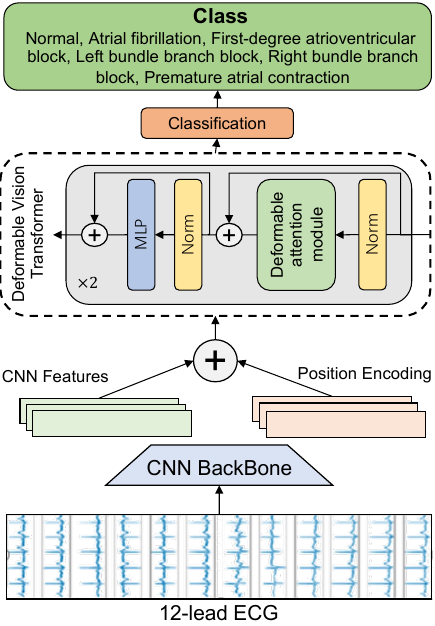} \\
	\caption{Overview of the GAI-based classification model, CNN-DVIT, proposed in [137]. It is able to take continuous 12-lead ECG signals as input and output the arrhythmia diagnosis result in an end-to-end manner. Specifically, the model is composed of three main components: (a) a CNN-based backbone for feature extraction from each lead; (b) a deformable attention transformer encoder module to combine the CNN-extracted features and the positional encoding; and (3) the classification layer to obtain the probability that each patient may have for each type of heart disease.}\label{DVIT}
\end{figure}

\begin{figure*}[!t]
	\centering
	\includegraphics[width=0.92\textwidth]{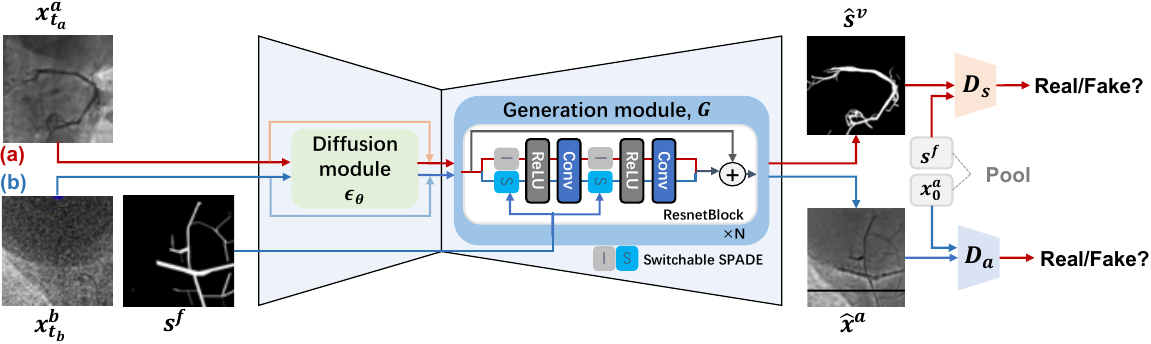} \\
	\caption{Overview of the diffusion model and GAN-based vessel segmentation method, DARL, proposed in [140]. In path (a), given a real noisy angiography image $x_{t_a}^a$, DARL estimates vessel segmentation mask $\hat{s}^v$. In path (b), given a real noisy background image $x_{t_b}^b$ and a vessel-like fractal mask $s^f$, DARL generates a synthetic angiography image $\hat{x}^a$.}\label{segmen}
\end{figure*}

The data classification is crucial for HDT services, such as early prevention and diagnosis of diseases. For example, when doctors remotely diagnose a patient through the patient's VT, classification models can quickly and automatically identify and classify the patient's diseases. Recent studies have shown the potential of the GAI-based methods in this field. For instance, Dhinagar et al. in \cite{301} investigated vision transformer (ViT) architecture for high-stakes neuroimaging downstream tasks, focusing on Alzheimer's disease classification based on 3D brain MRI. The authors evaluated the effects of different training strategies including pre-training, data augmentation and learning rate warm-ups followed by annealing, and emphasized the importance of these strategies in neuroimaging applications. Similarly, Dong et al. in \cite{302} proposed a ViT-based end-to-end multi-label arrhythmia classification model, called CNN-DVIT, for the 12-lead ECG with varied-length recordings, as shown in Fig. \ref{DVIT}. Specifically, CNN-DVIT was based on a combination of CNN with depthwise separable convolution, and a ViT architecture with deformable attention. Besides, the authors introduced the spatial pyramid pooling layer to accept varied-length ECG signals. Experimental results showed that CNN-DVIT outperformed the most recent transformer-based ECG classification methods \cite{382}.

Data segmentation play a crucial role in data analysis within the realm of HDT. Segmentation allows complex HDT data to be subdivided and analyzed to extract critical information. For instance, through image segmentation, the different regions and structures within an image can be separated, such as organs and tumors. This facilitates  quantitative analysis and measurements, such as measuring organ volumes and calculating tumor growth rates. This important information can be provided for HDT's subsequent services, such as surgical planning, treatment plans and diagnosis. GAI has been successfully implemented in this field.  It is worth note that , for data analysis in HDT, analyzing organs or other human structures from medical images is not a deterministic pixel-wise process, but underlies the assessment of the whole image or, on smaller scale, assessing the neighboring pixels' diversity. In this regard, Rahman et al. in \cite{304} leveraged the stochastic sampling step in the diffusion model to produce diverse and multiple masks. Specifically, the authors introduced a single diffusion model-based ambiguous segmentation network, called collectively intelligent medical diffusion (CIMD), that can generate multiple plausible annotations from a single input image. The CIMD utilized the noisy segmentation ground-truth masks concatenated to the original image to prevent the conventional diffusion process usage in the segmentation task from producing more resilient results, rather than arbitrary masks. The authors validated the CIMD in three different medical image modalities, namely CT, ultrasound and MRI. The experimental results showed that CIMD outperformed existing SOTA ambiguous segmentation networks in terms of accuracy while preserving naturally occurring variation. Similarly, Kim et al. in \cite{303} proposed a diffusion adversarial representation learning (DARL) model, which leveraged a denoising diffusion probabilistic mode with adversarial learning, for self-supervised vessel segmentation, aiming to diagnose vascular diseases and treatment planning, as shown in Fig. \ref{segmen}. Specifically, DARL model consisted of two main modules, where the diffusion module learned background image distribution, and the generation module generated vessel segmentation masks or synthetic angiograms using the proposed switchable spatially-adaptive denormalization algorithm. Experimental results showed that DARL model significantly outperforms existing unsupervised and self-supervised vessel segmentation methods.

In the context of HDT, anomaly detection for HDT data plays a critical role. Anomaly detection helps identify abnormal patterns or conditions in HDT data early, enabling proactive health alerts from HDT and timely intervention before a condition deteriorates. GAI has been successfully implemented in this field. For instance, Guo et al. in \cite{305} designed a cascade VAE-based anomaly detector (CVAD) for outlier detection in medical images. Specifically, with a focus on the generalizability of anomaly detector, CVAD combined latent representation at multiple scales, before being fed to a discriminator to distinguish the out-of-distribution (OOD) data from the in-distribution data. The reconstruction error and the OOD probability predicted by the binary discriminator were used to determine the anomalies. Extensive experiments on multiple intra- and inter-class OOD medical imaging datasets showed CVAD's effectiveness and generalizability. The GAN-based approaches are also successfully applied in anomaly detection. For instance, Nakao et al. in \cite{306} designed an auto-encoding GAN ($\alpha$-GAN) framework, which was a combination of a GAN and a VAE, for anomaly detection in chest radiographs. The experimental results showed that $\alpha$-GAN can correctly visualize various lesions including a lung mass, cardiomegaly, pleural effusion, bilateral hilar lymphadenopathy, and even dextrocardia. However, the VAE-based and GAN-based anomaly detection methods are often complicated to train or have difficulties to preserve fine details in the medical images \cite{300}. To this end, Wolleb et al. in \cite{300} proposed a weakly supervised anomaly detection method based on denoising diffusion implicit models (DDIMs)~\cite{du2023diffusion}. Specifically, the authors combined the deterministic iterative noising and denoising scheme with classifier guidance for image-to-image translation between diseased and healthy subjects. The authors applied this method on two different medical datasets, namely BRATS2020 brain tumor datasets and the CheXpert datasets \cite{383}. Experimental results showed that this method can preserves details of the input image unaffected by the disease while realistically representing the diseased part.

Prediction plays a critical role in data analysis, which can provide informed information for HDT services. For instance, the disease progression and drug response can be predicted in patient' VTs based on HDT data. With this, doctors can optimize the intervention and design the customized treatment plans for patients. GAI has been successfully implemented in this field. For instance, Allen et al. in \cite{249} proposed a $\beta$-VAE based HDT to forecast clinical measurement trajectory in patients who went on to experience an ischemic stroke. Specifically, the $\beta$-VAE based HDT model was used to generate possible next steps in a patient's disease progression. The model was trained on data extracted from the medical information mart for intensive care-IV database \cite{384}. The database contains 1216 patients with useable trajectories that experienced ischemic stroke. Experimental results demonstrated that the model can accurately forecast the progression of relevant clinical measurements in patients at risk of ischemic stroke, which was virtually indistinguishable from real patient data. Additionally, transformer-based approaches have also been used in prediction of disease progression. For instance, Hu et al. in \cite{307} applied four BERT-based models, including the original BERT \cite{308}, BioBERT \cite{309}, RoBERTa \cite{310}, and DistilBERT \cite{311}, to ophthalmology clinical text from EHR, predicting the progression of glaucoma. Experiment results showed that these four BERT-based models outperformed clinical predictions by an ophthalmologist's review of the same clinical information. The authors qualitatively evaluated the BERT-based models by performing explainability studies based on the self-attention mechanisms of the BERT-models. Based on this, the authors can evaluate what types of words were most important for predictions.

In summary, GAI is a key enabling technology for data analysis in HDT, analyzing the HDT data for classification, segmentation, anomaly detection and prediction.

\subsection{Promoting the GAI-driven HDT}
In order to successfully integrate GAI-driven HDT technology into IoT-healthcare, it is vital to involve both patients and healthcare professionals in the process. Here we provide a few strategies for engaging them in the development and adoption of GAI-driven HDT.

\emph{1) Educational workshops and training}: Start by building knowledge about GAI-driven HDT among healthcare professionals. Educating them on the benefits and limitations, explaining how it can be effectively used, discussing privacy concerns, and any possible adverse effects, can enable healthcare professionals to use it more effectively and also communicate these aspects convincingly to the patients.

\emph{2) Collaborative development}: Involve healthcare professionals in the development process of these applications to ensure they align with current workflow and improve healthcare delivery. In turn, the developer should also do navigation for understanding the demands and concerns of patients. Their first-hand insights are invaluable to make the technology even more helpful and user-friendly.

\emph{3) Ethical and privacy concerns}: To mitigate the potential ethical and privacy issues, there needs to be a solid foundation of transparency. Ethical AI is imperative to be considered. There needs to be a solid foundation of transparency. The GAI-driven HDT technology providers should be able to provide clear explanations of its processes and decisions to both patients healthcare professionals. This means explicit consent should be obtained before any personal health data is collected or used. The GAI-driven HDT should be trained and tested to ensure that it does not exhibit any bias, ensuring fairness and equity in healthcare provisioning. Additionally, accountability and oversight are needed to ensure the individuals or the entities responsible for the GAI-driven HDT making its corresponding decisions accountable. Besides, the strategies for ensuring data privacy and security are crucial. Technologies such as blockchain \cite{409}, federated learning \cite{393}, differential privacy \cite{66}, secure multi-party computation \cite{407}, and physical layer security aware wireless communications \cite{399, 400, 401}, can be designed and implemented for GAI-driven HDT to prevent unauthorized access and data breaches.

\section{GAI-driven HDT in IoT-Healthcare Applications} \label{SE4}
The remarkable capabilities of GAI-driven HDT in a wide range of IoT-healthcare applications, including personalized health monitoring and diagnosis, prescription, and rehabilitation. In this section, we will delve into the details of these IoT-healthcare applications, exploring how GAI-driven HDT is revolutionizing each of them.

\subsection{Personalized Health Monitoring and Diagnosis}

\begin{figure*}[!t]
	\centering
	\includegraphics[width=0.98\textwidth]{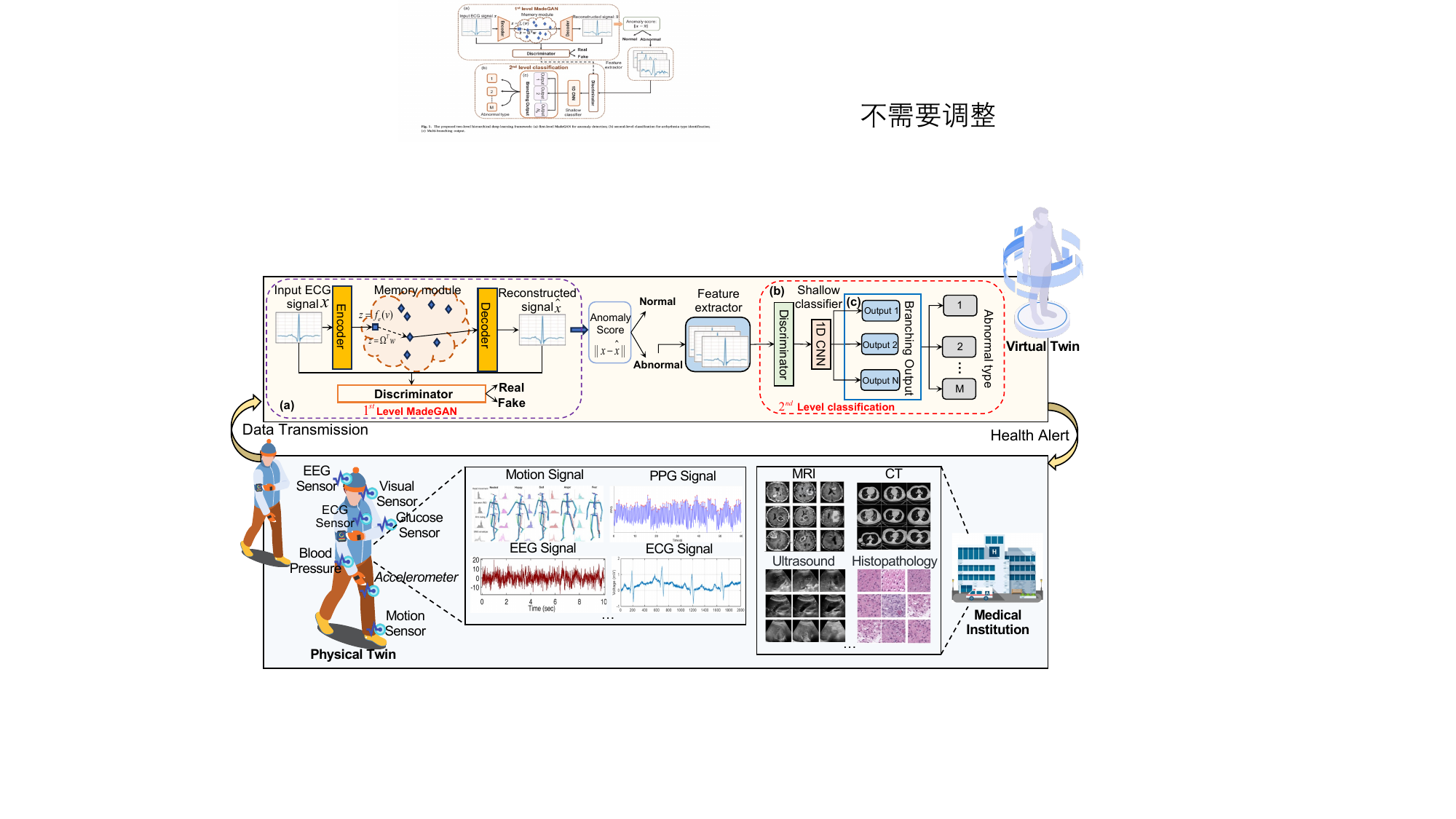} \\
	\caption{The application of GAI-driven HDT in the personalized health monitoring and diagnosis, where the GAN-based approach, proposed in [158] is used as an example. The collected ECG signals from the physical twin are transmitted to the GAI-driven HDT for ECG signal analysis. Specifically, the GAN-based approach is a two-level hierarchical deep learning framework: (a) first-level MadeGAN for anomaly detection; (b) second-level classification for arrhythmia type identification; (c) Multi-branching output. If there are abnormities in collected data, the virtual twin will initiate health alerts to either the physical twin or medical institution for interventions.}\label{healmoni}
\end{figure*}

The GAI for HDT can be applied in personalized health monitoring and diagnosis. By leveraging the robust data analysis capabilities, GAI can analyze collected patient data in HDT, enabling anomaly detection for personalized health monitoring and diagnosis, as shown in Fig. \ref{healmoni}. For example, a VT continuously receives data streams from its corresponding PT, namely the patient, encompassing vital metrics like heart rate, blood pressure, and ECG signals, etc. Utilizing GAI, the VT can analyze this data, identifying deviations from expected norms, such as cardiac irregularities or other health concerns. Detected anomalies trigger alerts to the patient and healthcare providers, signaling potential irregularities or cardiac issues. Consequently, GAI for HDT is a personalized health monitoring system that delivers timely health warnings.

GAI has been successfully applied in anomaly detection from the collected patient data in HDT. For example, Wang et al. in \cite{341} proposed a GAN-based approach for ECG signal analysis, achieving automatic cardiac diagnosis, as shown in Fig. \ref{healmoni}. Specifically, the proposed approach involved two-level hierarchical deep learning framework with GAN. The first-level model was composed of a memory-augmented deep autoencoder with GAN (MadeGAN), which aimed to differentiate abnormal signals from normal ECGs for anomaly detection. The second-level learning aimed at robust multi-class classification for different arrhythmia identification, which was achieved by integrating the transfer learning technique to transfer knowledge from the first-level learning with the multi-branching architecture to handle the data-lacking and imbalanced data issues. The performance of the GAN-based anomaly detection was evaluated by two global performance metrics, i.e., area under receiver-operating-characteristic curve (AUROC), and area under precision-recall curve (AUPRC). The ROC characterizes the relationship between false positive rate (FPR) and true positive rate (TPR). Then, AUROC is a score quantifying the general prediction performance across all thresholds. Meanwhile, the AUPRC evaluates the general relationship between recall and precision. In the GAN-based anomaly detection, the performance of the proposed MadeGAN was superior than traditional methods, i.e., AutoEncoder (AE) and BeatGAN \cite{newstart_425}. Specifically, MadeGAN achieved the best performance with AUROC and AUPRC scores of 0.954 and 0.936. In addition, MadeGAN improved AUROC by 5.3\% compared with AE (0.906), and by 1.4\% compared with BeatGAN (0.941). Furthermore, the improvement of AUPRC is more significant, with an increase of 5.6\% and 1.5\% compared with AE (0.886) and BeatGAN (0.922), respectively.

In addition, the industry is also walking on this way, aiming to facilitate the implementation of GAI-driven HDT. The global leader in GAI for pathology, Paige, leverages large and diverse dataset in cancer diagnostics to develop their GAI-based products for cancer pathology detection and analysis \cite{391}. Their GAI-based products has been independently validated and clinically utilized by hospitals and labs around the world, and the reduction in cancer detection errors reaches 70\% \cite{390}, while the reduction in time to diagnosis is 65.5\% \cite{389}. It is expected that such powerful tools can be integrated into HDT for personalized health monitoring and diagnosis.

Thus, GAI-driven HDT holds the promise of delivering precise, timely, and proactive personalized health monitoring and diagnosis, enhancing patient care and overall well-being.

\subsection{Personalized Prescription}

\begin{figure*}[!t]
	\centering
	\includegraphics[width=0.89\textwidth]{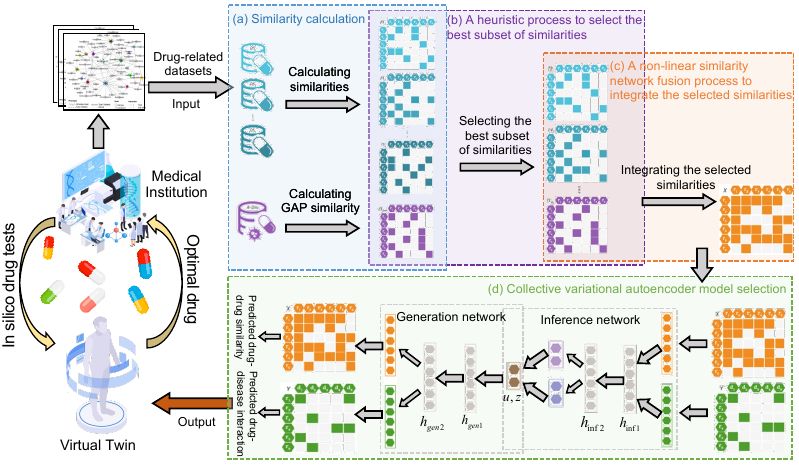} \\
	\caption{The application of GAI-driven HDT in the personalized prescription, where the GAI-based approach, SNF–CVAE method, proposed in [163] is used as an example: (a) Calculating drug similarity matrices using the drug-related datasets obtaining from the medical institution. (b) Applying a heuristic process to select the most informative, least redundant subset of drug similarity matrices. (c) Applying a non-linear similarity network fusion process to integrate the selected similarity matrices into a comprehensive drug similarity matrix. (d) Training a collective VAE model with drug similarity information and drug–disease interactions to predict novel drug–disease interactions. Finally, the output is used to choose the optimal drug for the patient.}\label{prescri}
\end{figure*}

The GAI-driven HDT can be applied in the personalized prescription. GAI empowers the HDT to predict and simulate treatment outcomes, enabling personalized prescription, as shown in Fig. \ref{prescri}. For instance, considering an Alzheimer's patient's VT, doctors can virtually test various candidate drugs on the VT. Then, leveraging GAI within the VT, the interactions of each drug with the patient's Alzheimer's disease can be predicted and simulated. Upon deriving insights from the GAI-driven HDT, doctors can pinpoint the most effective drug tailored to the patient's specific health profile.

GAI has been successfully applied in treatment outcome predictions and simulations inside the HDT for personalized prescriptions. For instance, Jarada et al. in \cite{346} proposed a VAE-based approach, called similarity network fusion-collective VAE (SNF-CVAE), for predicting drug-disease interactions, as shown in Fig. \ref{prescri}. Specifically, SNF-CVAE integrated similarity measures, similarity selection, SNF, and CVAE to conduct a non-linear analysis and improve the drug-disease interaction prediction accuracy. For evaluating the performance of SNF-CVAE in drug-disease interactions, authors introduced five evaluation metrics, i.e., precision, recall, F1-score, area under curve-receiver operator characteristic (AUC-ROC), and area under curve-precision-recall (AUC-PR) \cite{427}. Specifically, Precision $= TP/(TP+FP)$, Recall $= TP/(TP+FN)$, and F1-score $= 2 * (Precision * Recall)/(Precision + Recall)$. $TP$ was predicted and known drug-disease interactions, $TN$ was unpredicted and unknown drug–disease interactions, $FP$ was predicted and unknown drug–disease interactions, and $FN$ was unpredicted and known drug–disease interactions. Additionally, AUC-ROC can be described as plotting the $TP$ against $FP$, while AUC-PR can be described as a trade-off between Precision and Recall. The simulation results showed that SNF-CVAE achieved significantly higher accuracy with Precision = 0.902, Recall = 0.883, F1-score = 0.893, AUC-ROC = 0.958 and AUC-PR = 0.970, compared the traditional methods, Pareto dominance and collaborative filtering-based drug-disease prediction \cite{428} with Precision = 0.487, Recall = 0.085, F1-score = 0.144, AUC-ROC = 0.542 and AUC-PR = 0.225. 

Beyond the academic successes, the industry has also achieved some milestones in this area. The pioneer in utilizing GAI for drug discovery and development, Insilico Medicine, had developed a GAI powered commercially available platform, Pharma.AI \cite{239}. Their lead drug, INS018\_055, was discovered and designed using Pharma.AI, which was designed to treat the devastating lung condition idiopathic pulmonary fibrosis (IPF) \cite{388}. IPF is a disease marked by excessive scarring of the lungs that affects 5 million people worldwide and has few treatment options. INS018\_055 was nominated as a preclinical candidate – a record 18 months from target discovery — and quickly accelerated into further trials, from first-in-human, to Phase I trials with healthy volunteers, where the drug showed excellent results, delivering a favorable pharmacokinetic and safety profile. Now (i.e., March 2024), the drug is currently in Phase 2a trials in both the U.S. and China with IPF patients – less than 4 years from discovery. Thus, compared to traditional drug discovery methods, which typically takes from 10 to 15 years to develop a drug, the GAI-driven methods presented here has significantly shorten the process. Promisingly, Pharma.AI can be integrated into the development of GAI-driven HDT, which can acts the human digital testbed by simulating the drug-disease interactions.

The continuous advancement of GAI-driven HDT can offer the potential for more diverse and accurate predictions and simulations of treatment outcomes. This capability facilitates more versatile personalized prescriptions, ultimately optimizing treatment outcomes.

\subsection{Personalized Rehabilitation}

\begin{figure*}[!t]
	\centering
	\includegraphics[width=0.98\textwidth]{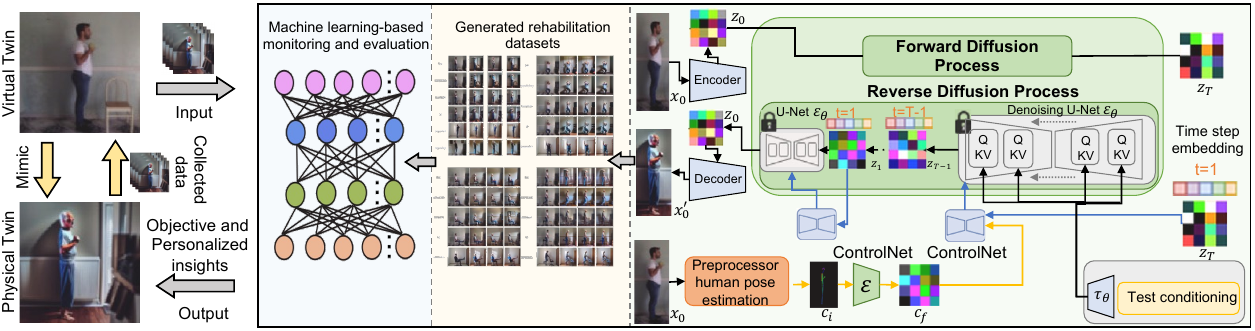} \\
	\caption{The application of GAI-driven HDT in the personalized rehabilitation, where the GAI-based approach, pose-guided condition diffusion model, proposed in [168] is used as an example. The physical twin mimics the motions of virtual twin, and the motions of the physical twin are collected and input into the machine learning model for obtaining the physical twin's objective and personalized insights. The training datasets of the machine learning model are enriched by the pose-guided condition diffusion model, which is based on a latent diffusion model combined with the ControlNet.}\label{reha}
\end{figure*}

The GAI for HDT can be applied in the personalized rehabilitation training. Based on the built patient's VT, the doctor or physical therapist can simulate different rehabilitation training scenarios by controlling the VT's target posture. For example, if the patient needs balance training, the doctor can adjust the VT’s posture to create a personalized unstable state that requires balance subject to the patient's profile. Then, the patient, namely the PT, observes the VT’s movements and attempts to mimic its posture to improve their balance. Then, the IoT devices are used to collect the patient's motion data, and transmit to the VT for monitoring and evaluation. Then, the VT serves as virtual ``coach'' providing real-time feedback and guidance to help the patient posture correction and actively progress through their rehabilitation training. 

During this personalized rehabilitation training, machine learning-based approaches are usually utilized to enhance the monitoring and evaluation process in the VT, providing objective and personalized insights. This method relies on substantial high-quality data to obtain a robust and accurate machine learning model. This data should encompass a diverse range of exercises and rehabilitation contexts, as well as holistic patient profiles. However, acquiring such data poses significant obstacles, such as data availability and privacy concerns, especially in individual-level. To this end, GAI can be used to generate synthetic data in the context of personalized rehabilitation training for enhancing the performance of the virtual ``coach'' role of VT, as shown in Fig. \ref{reha}. For instance, Mennella et al. in \cite{338} proposed a diffusion model-based approach, pose-guided condition diffusion model, to generate synthetic data that mimicked realistic-looking human movements in a rehabilitation context. Specifically, the data generation framework used the latent diffusion model in combination with ControlNet \cite{339}, and was trained on a pre-labeled dataset of 6 rehabilitation exercises. To evaluate the performance of the proposed approach, authors adopted qualitative and quantitative evaluation methods. As a rehabilitation specialist with extensive expertise in biomechanical and motion analysis, the first author of this paper undertook the assessment of the generated images for qualitative evaluation. As for quantitative evaluation, authors adopted the Inception score (IS), the Frechet Inception distance (FID), the Structural similarity index (SSMI), and the detection score for person-class (DS-person). These metrics provided a comprehensive evaluation of the fidelity, quality, and visual resemblance of the generated images to the ground truth images. Experiment results showed that the synthetic datasets achieved high-quality performance, i.e., IS = 162.94, FID = 299.06, SSMI = 0.545, DS-person = 0.789.

These synthetic datasets can be tailored to target various aspects of rehabilitation, including postural correction, joint mobility, balance training, and functional movements. Additionally, it can also be customized according to patient profiles, achieving personalized rehabilitation training. In summary, these GAI-based data augmentation methods are significant for HDT in rehabilitation training by enhancing the monitoring and evaluation process.

\section{Future Research Directions} \label{SE5}
The GAI-driven HDT in IoT-healthcare is revolutionizing the healthcare industry, as surveyed above. This exciting field is nascent, and therefore, some critical issues remain unexplored and are of great importance.

\subsection{Deriving Energy-Efficient  Scheme}
Implementing such a large model, GAI-driven HDT, requires substantial requirements in computation, communication, and storage resources due to the high parametric complexity of GAI-driven HDT and the necessity for vast datasets. For instance, 10,000 graphical processing unit (GPU) cards were employed to run an HDT brain \cite{185, 186}. It will certainly result in large energy consumption, and generate significant carbon dioxide emissions, jeopardizing sustainability. Thus, the energy-efficient scheme for the implementation of GAI-driven HDT is imperative. Techniques such as pruning, knowledge distillation, quantization, and green learning, have been proposed to address this issue. However, these solutions come at the cost of sacrificing the accuracy of GAI-driven HDT, which is a critical issue, especially in IoT-healthcare. Therefore, it should carefully balance accuracy and sustainability in energy-efficient GAI-driven HDT.

\subsection{Designing Human-centric Evaluation Metrics}
Human-centric metric design for evaluating the performance of GAI-driven HDT in IoT-healthcare is inherently challenging. It should encompass the accuracy and reliability of the GAI generated content compared to the PTs' real health data. This involves defining parameters to evaluate the fidelity of the VTs' behavior, responses, and predictive capabilities in the context of the IoT-healthcare enabled by GAI.
One of potential solution is the integration of GAI evaluation metrics, such as inception score and Frechet inception distance, and healthcare knowledge. Based on these, the continual evaluation and refinement of the GAI-driven HDT can ultimately enhance the performance in IoT-healthcare.

\subsection{Accelerating Real-time Responsiveness}
The slow generation process of GAI models poses significant challenges for developing GAI-driven HDT in IoT-healthcare. This delay hampers real-time applications, such as immersive interactions with VTs, and limits the responsiveness necessary for instant healthcare interventions, such as timely diagnostics or immediate treatment recommendations in critical conditions. Thus, accelerating real-time responsiveness of GAI-driven HDT is imperative to ensure timely and effective healthcare outcomes. One potential solution is refining and optimizing GAI algorithms to expedite the generation process. Advanced parallel computing techniques, such as leveraging high-performance computing or distributed computing frameworks, can significantly accelerate model training and generation, mitigating the latency and enhancing agility in responding to critical healthcare needs.

\subsection{Seamless Integration of GAI-driven HDT with Existing Healthcare IT Ecosystems}

The integration of GAI-driven HDT into existing healthcare IT ecosystems is a fundamental requirement for the successful practical implementation. In this process, numerous many interoperability issues, such as efficient data exchange and effective data integration between these two types of systems. Standardized application programming interfaces (APIs) and web services  is a practical solution to these challenges. In this regard, one significant integration issue to overcome is the existence of various data structures and standards within these two types of systems. To confront this complexity, deploying uniform data formats, such as health level seven (HL7) and fast healthcare interoperability resources (FHIR), is advisable. These formats are widely recognizable within global healthcare IT infrastructures, and their use in GAI-driven HDT can aid it in communicating effectively with existing healthcare IT infrastructures, allowing a streamlined process of data exchange and integration. However, the adoption of HL7 and FHIR in GAI-driven HDT would arise many challenges. For example, both them are complex data standards, with a broad range of elements and structures, which will pose a challenge for the GAI-driven HDT in terms of data processing and understanding. Additionally, mapping the data from the GAI-driven HDT to these formats is another challenging task due to varying data types and intricate data relationships. Tackling these integration and interoperability issues would provide a clearer roadmap for the practical implementation of GAI- driven HDT in IoT-healthcare.

\section{Conclusions} \label{SE6}
In this survey, we have shed light on implementing GAI-driven HDT in IoT-healthcare. Specifically, we have introduced IoT-healthcare, and envisioned the potential of utilizing GAI-driven HDT. Then, we have reviewed the fundamentals of HDT and GAI and illustrated the framework of GAI-driven HDT. After that, we have analyzed the implementation of GAI-driven HDT:

1) GAI can assist the data acquisition in HDT, by generating diverse and highly realistic synthetic data, including synthetic physiological, medical imaging and motion data.

2) GAI can offer efficient communication schemes for HDT, including GAI-aid semantic communication and GAI-aid cross-modal communication.

3) GAI can be used in data management of HDT by generating the missed data for data imputation, and clean data for data denoising. Additionally, GAI can produce synthesized yet authentic HDT data largely, preserving the original data distribution, whilst ensuring data security and privacy.

4) GAI can empower digital modeling in HDT, including digital modeling of human cells, tissues and organs.

5) GAI can facilitate data analysis in HDT, including classification, segmentation, anomaly detection and prediction.

Based on this, we have further presented the IoT-healthcare applications of GAI-driven HDT, such as personalized health monitoring and diagnosis, personalized prescription, and personalized rehabilitation. Finally, we have outlined some future research directions.

\section*{Acknowledgments}

This work was supported by the State Key Laboratory of Massive Personalized Customization System and Technology under grant No. H\&C-MPC-2023-04-01, and the National Research Foundation, Singapore, and Infocomm Media Development Authority under its Future Communications Research \& Development Programme, Defence Science Organisation (DSO) National Laboratories under the AI Singapore Programme (AISG Award No: AISG2-RP-2020-019 and FCP-ASTAR-TG-2022-003), and Singapore Ministry of Education (MOE) Tier 1 (RG87/22), and the National Natural Science Foundation of China (NSFC) under Grants No. 62102099 and No. U22A2054, and Postgraduate Research \& Practice Innovation Program of Jiangsu Province No.KYCX24\_0596.

\bibliographystyle{IEEEtran}
\bibliography{ref}

\begin{thebibliography}{100}
\providecommand{\url}[1]{#1}
\csname url@samestyle\endcsname
\providecommand{\newblock}{\relax}
\providecommand{\bibinfo}[2]{#2}
\providecommand{\BIBentrySTDinterwordspacing}{\spaceskip=0pt\relax}
\providecommand{\BIBentryALTinterwordstretchfactor}{4}
\providecommand{\BIBentryALTinterwordspacing}{\spaceskip=\fontdimen2\font plus
\BIBentryALTinterwordstretchfactor\fontdimen3\font minus
  \fontdimen4\font\relax}
\providecommand{\BIBforeignlanguage}[2]{{%
\expandafter\ifx\csname l@#1\endcsname\relax
\typeout{** WARNING: IEEEtran.bst: No hyphenation pattern has been}%
\typeout{** loaded for the language `#1'. Using the pattern for}%
\typeout{** the default language instead.}%
\else
\language=\csname l@#1\endcsname
\fi
#2}}
\providecommand{\BIBdecl}{\relax}
\BIBdecl

\bibitem{415}
M.~M. Islam, S.~Nooruddin, F.~Karray \emph{et~al.}, ``Internet of things:
  Device capabilities, architectures, protocols, and smart applications in
  healthcare domain,'' \emph{IEEE Internet Things J.}, vol.~10, no.~4, pp.
  3611--3641, 2023.

\bibitem{416}
M.~N. Bhuiyan, M.~M. Rahman, M.~M. Billah \emph{et~al.}, ``Internet of things
  {(IoT)}: A review of its enabling technologies in healthcare applications,
  standards protocols, security, and market opportunities,'' \emph{IEEE
  Internet Things J.}, vol.~8, no.~13, pp. 10\,474--10\,498, 2021.

\bibitem{349}
S.~Selvaraj and S.~Sundaravaradhan, ``Challenges and opportunities in {IoT}
  healthcare systems: a systematic review,'' \emph{SN Appl. Sci.}, vol.~2,
  no.~1, p. 139, 2020.

\bibitem{350}
M.~H. Kashani, M.~Madanipour, M.~Nikravan \emph{et~al.}, ``A systematic review
  of {IoT} in healthcare: Applications, techniques, and trends,'' \emph{J.
  Netw. Comput. Appl.}, vol. 192, p. 103164, 2021.

\bibitem{356}
G.~H. Hub, ``Global heart hub,'' \url{https://globalhearthub.org}, 2023.

\bibitem{352}
H.~Habibzadeh, K.~Dinesh, O.~R. Shishvan \emph{et~al.}, ``A survey of
  healthcare internet of things {(HIoT)}: A clinical perspective,'' \emph{IEEE
  Internet Things J.}, vol.~7, no.~1, pp. 53--71, 2020.

\bibitem{351}
Y.~Yang, H.~Wang, R.~Jiang \emph{et~al.}, ``A review of {IoT}-enabled mobile
  healthcare: technologies, challenges, and future trends,'' \emph{IEEE
  Internet Things J.}, vol.~9, no.~12, pp. 9478--9502, 2022.

\bibitem{418}
V.~Hayyolalam, M.~Aloqaily \emph{et~al.}, ``Edge-assisted solutions for
  {IoT}-based connected healthcare systems: A literature review,'' \emph{IEEE
  Internet Things J.}, vol.~9, no.~12, pp. 9419--9443, 2022.

\bibitem{420}
C.~Advisory, ``The history and creation of the digital twin concept,''
  \url{https://www.challenge.org/insights/digital- twin-history/}, 2021.

\bibitem{419}
Y.~Wu, K.~Zhang, and Y.~Zhang, ``Digital twin networks: A survey,'' \emph{IEEE
  Internet Things J.}, vol.~8, no.~18, pp. 13\,789--13\,804, 2021.

\bibitem{054}
T.~C. S. L. A.~H. of~the Informatics~Institute, ``Ecosystem for digital twins
  in healthcare,'' \url{https://www.edith-csa.eu/}, 2023.

\bibitem{3}
B.~Wang, H.~Zhou, X.~Li \emph{et~al.}, ``Human digital twin in the context of
  {Industry} 5.0,'' \emph{Robotics and Computer-Integrated Manufacturing},
  vol.~85, p. 102626, 2024.

\bibitem{4}
J.~Chen, C.~Yi, S.~D. Okegbile \emph{et~al.}, ``Networking architecture and key
  supporting technologies for human digital twin in personalized healthcare: A
  comprehensive survey,'' \emph{IEEE Commun. Surv. Tutor.}, vol.~26, no.~1, pp.
  706--746, 2024.

\bibitem{5}
S.~D. Okegbile, J.~Cai, C.~Yi \emph{et~al.}, ``Human digital twin for
  personalized healthcare: Vision, architecture and future directions,''
  \emph{IEEE Netw.}, 2022.

\bibitem{6}
A.~E. Co{\c{s}}gun, ``Digital twin and human digital twin for practical
  implementation in industry 5.0,'' in \emph{Global Perspectives on Robotics
  and Autonomous Systems: Development and Applications}.\hskip 1em plus 0.5em
  minus 0.4em\relax IGI Global, 2023, pp. 168--183.

\bibitem{18}
B.~Bj{\"o}rnsson, C.~Borrebaeck, N.~Elander \emph{et~al.}, ``Digital twins to
  personalize medicine,'' \emph{Genome Med.}, vol.~12, pp. 1--4, 2020.

\bibitem{235}
Google, ``Med-palm,'' \url{ https://sites.research.google/med-palm/ }, 2023.

\bibitem{8}
J.~Chen, C.~Yi, H.~Du \emph{et~al.}, ``A revolution of personalized healthcare:
  Enabling human digital twin with mobile {AIGC},'' \emph{IEEE Netw.}, 2024.

\bibitem{du2023beyond}
H.~Du, R.~Zhang, Y.~Liu, J.~Wang, Y.~Lin, Z.~Li, D.~Niyato, J.~Kang, Z.~Xiong,
  S.~Cui \emph{et~al.}, ``Beyond deep reinforcement learning: A tutorial on
  generative diffusion models in network optimization,'' \emph{arXiv preprint
  arXiv:2308.05384}, 2023.

\bibitem{10}
S.~Bond-Taylor, A.~Leach, Y.~Long \emph{et~al.}, ``Deep generative modelling: A
  comparative review of {VAEs, GANs}, normalizing flows, energy-based and
  autoregressive models,'' \emph{IEEE Trans. Pattern Anal. Mach. Intell.},
  vol.~44, no.~11, pp. 7327--7347, 2022.

\bibitem{421}
D.~Moher, A.~Liberati, J.~Tetzlaff \emph{et~al.}, ``Preferred reporting items
  for systematic reviews and meta-analyses: the {PRISMA} statement,''
  \emph{Ann. Intern. Med.}, vol. 151, no.~4, pp. 264--269, 2009.

\bibitem{s1}
Y.~Lin, L.~Chen, A.~Ali \emph{et~al.}, ``Human digital twin: A survey,''
  \emph{arXiv preprint arXiv:2212.05937}, 2022.

\bibitem{s2}
M.~Xu, H.~Du, D.~Niyato \emph{et~al.}, ``Unleashing the power of edge-cloud
  generative {AI} in mobile networks: A survey of aigc services,'' \emph{arXiv
  preprint arXiv:2303.16129}, 2023.

\bibitem{s3}
B.~Wang, H.~Zhou, X.~Li \emph{et~al.}, ``Human digital twin in the context of
  industry 5.0,'' \emph{Robot. Comput.-Integr. Manuf.}, vol.~85, p. 102626,
  2024.

\bibitem{s4}
J.~Chen, C.~Yi, S.~D. Okegbile \emph{et~al.}, ``Networking architecture and key
  supporting technologies for human digital twin in personalized healthcare: A
  comprehensive survey,'' \emph{IEEE Commun. Surv. Tutor.}, 2023.

\bibitem{s5}
P.~Pataranutaporn, V.~Danry, J.~Leong \emph{et~al.}, ``{AI}-generated
  characters for supporting personalized learning and well-being,'' \emph{Nat.
  Mach. Intell.}, vol.~3, no.~12, pp. 1013--1022, 2021.

\bibitem{s6}
H.~Pascual, X.~M. Bruin, A.~Alonso \emph{et~al.}, ``A systematic review on
  human modeling: Digging into human digital twin implementations,''
  \emph{arXiv preprint arXiv:2302.03593}, 2023.

\bibitem{353}
T.~Sun, X.~He, and Z.~Li, ``Digital twin in healthcare: Recent updates and
  challenges,'' \emph{Digit. Health}, vol.~9, p. 20552076221149651, 2023.

\bibitem{354}
M.~AlAmir and M.~AlGhamdi, ``The role of generative adversarial network in
  medical image analysis: An in-depth survey,'' \emph{ACM Comput. Surv.},
  vol.~55, no.~5, pp. 1--36, 2022.

\bibitem{355}
Y.~Shokrollahi, S.~Yarmohammadtoosky, M.~M. Nikahd \emph{et~al.}, ``A
  comprehensive review of generative ai in healthcare,'' \emph{arXiv preprint
  arXiv:2310.00795}, 2023.

\bibitem{13}
\BIBentryALTinterwordspacing
E.~Research, ``Digital human avatar market, by product type (interactive
  digital human avatar and non-interactive digital human avatar), by industry
  verticals (gaming, retail, it \& telecommunications, education, and others),
  and by region forecast to 2032,'' \emph{Emergen Research}, 2023. [Online].
  Available:
  \url{https://www.emergenresearch.com/industry-report/digital-human-avatar-market}
\BIBentrySTDinterwordspacing

\bibitem{7}
H.~Lonsdale, G.~M. Gray, L.~M. Ahumada \emph{et~al.}, ``The perioperative human
  digital twin,'' \emph{Anesth. Analg.}, vol. 134, no.~4, pp. 885--892, 2022.

\bibitem{22}
F.~Tao, H.~Zhang, A.~Liu \emph{et~al.}, ``Digital twin in industry:
  State-of-the-art,'' \emph{IEEE Trans. Industr. Inform.}, vol.~15, no.~4, pp.
  2405--2415, 2019.

\bibitem{21}
S.~Jang, J.~Jeong, J.~Lee \emph{et~al.}, ``Digital twin for intelligent
  network: Data lifecycle, digital replication, and ai-based optimizations,''
  \emph{IEEE Commun.}, pp. 1--7, 2023.

\bibitem{20}
Z.~Hu, S.~Lou, Y.~Xing \emph{et~al.}, ``Review and perspectives on driver
  digital twin and its enabling technologies for intelligent vehicles,''
  \emph{IEEE Trans. Intell. Veh.}, vol.~7, no.~3, pp. 417--440, 2022.

\bibitem{23}
A.~Bilberg and A.~A. Malik, ``Digital twin driven human--robot collaborative
  assembly,'' \emph{CIRP annals}, vol.~68, no.~1, pp. 499--502, 2019.

\bibitem{24}
B.~M. Rosa and G.~Z. Yang, ``A flexible wearable device for measurement of
  cardiac, electrodermal, and motion parameters in mental healthcare
  applications,'' \emph{IEEE J. Biomed. Health Inform.}, vol.~23, no.~6, pp.
  2276--2285, 2019.

\bibitem{25}
R.~Ferdousi, F.~Laamarti, M.~A. Hossain \emph{et~al.}, ``Digital twins for
  well-being: an overview,'' \emph{Digital Twin}, vol.~1, p.~7, 2022.

\bibitem{26}
Y.~Lin, L.~Chen, A.~Ali \emph{et~al.}, ``Human digital twin: A survey,''
  \emph{arXiv preprint arXiv:2212.05937}, 2022.

\bibitem{27}
H.~Y. Zhu, N.~Q. Hieu, D.~T. Hoang \emph{et~al.}, ``A human-centric metaverse
  enabled by brain-computer interface: A survey,'' \emph{arXiv preprint
  arXiv:2309.01848}, 2023.

\bibitem{11}
S.~Sai, A.~Gaur, R.~Sai \emph{et~al.}, ``Generative {AI} for transformative
  healthcare: A comprehensive study of emerging models, applications, case
  studies, and limitations,'' \emph{IEEE Access}, vol.~12, pp.
  31\,078--31\,106, 2024.

\bibitem{429}
G.~O. Ghosheh, J.~Li, and T.~Zhu, ``A survey of generative adversarial networks
  for synthesizing structured electronic health records,'' \emph{ACM Comput.
  Surv.}, vol.~56, no.~6, pp. 1--34, 2024.

\bibitem{430}
D.~Papadopoulos and V.~D. Karalis, ``Variational autoencoders for data
  augmentation in clinical studies,'' \emph{Appl. Sci.}, vol.~13, no.~15, p.
  8793, 2023.

\bibitem{431}
K.~Denecke, R.~May, and O.~Rivera-Romero, ``Transformer models in healthcare: A
  survey and thematic analysis of potentials, shortcomings and risks,''
  \emph{J. Med. Syst.}, vol.~48, no.~1, p.~23, 2024.

\bibitem{83}
A.~Kazerouni \emph{et~al.}, ``Diffusion models in medical imaging: A
  comprehensive survey,'' \emph{Med. Image Anal.}, p. 102846, 2023.

\bibitem{035}
A.~Ferreira, J.~Li, K.~L. Pomykala \emph{et~al.}, ``{GAN}-based generation of
  realistic {3D} volumetric data: A systematic review and taxonomy,''
  \emph{Med. Image Anal.}, p. 103100, 2024.

\bibitem{313}
T.~Golany, K.~Radinsky, and D.~Freedman, ``{SimGANs}: Simulator-based
  generative adversarial networks for {ECG} synthesis to improve deep {ECG}
  classification,'' in \emph{Proc. ICML}.\hskip 1em plus 0.5em minus
  0.4em\relax PMLR, 2020, pp. 3597--3606.

\bibitem{286}
F.~Lau, T.~Hendriks, J.~Lieman-Sifry \emph{et~al.}, ``{ScarGAN}: chained
  generative adversarial networks to simulate pathological tissue on
  cardiovascular {MR} scans,'' in \emph{Proc. DLMIA}.\hskip 1em plus 0.5em
  minus 0.4em\relax Springer, 2018, pp. 343--350.

\bibitem{036}
Y.~Wang, C.~Li, and Z.~Wang, ``Advancing precision medicine: {VAE} enhanced
  predictions of pancreatic cancer patient survival in local hospital,''
  \emph{IEEE Access}, vol.~12, pp. 3428--3436, 2024.

\bibitem{325}
A.~Dittadi, S.~Dziadzio, D.~Cosker, and ohters, ``Full-body motion from a
  single head-mounted device: Generating {SMPL} poses from partial
  observations,'' in \emph{Proc. IEEE/CVF ICCV}, 2021, pp. 11\,687--11\,697.

\bibitem{433}
E.~V. Mascar{\'o}, H.~Ahn, and D.~Lee, ``A unified masked autoencoder with
  patchified skeletons for motion synthesis,'' in \emph{Proc. AAAI}, vol.~38,
  no.~6, 2024, pp. 5261--5269.

\bibitem{249}
A.~Allen, A.~Siefkas, E.~Pellegrini \emph{et~al.}, ``A digital twins machine
  learning model for forecasting disease progression in stroke patients,''
  \emph{Appl. Sci.}, vol.~11, no.~12, p. 5576, 2021.

\bibitem{034}
C.~Zhang, L.~Liu, J.~Dai \emph{et~al.}, ``Xtransct: ultra-fast volumetric ct
  reconstruction using two orthogonal x-ray projections for image-guided
  radiation therapy via a transformer network,'' \emph{Phys. Med. Biol.},
  vol.~69, no.~8, p. 085010, 2024.

\bibitem{233}
Microsoft and Epic, ``Alphasense,'' \url{https://research.alpha-sense.com},
  2023.

\bibitem{278}
H.~Cui, C.~Wang, H.~Maan \emph{et~al.}, ``{scGPT}: Towards building a
  foundation model for single-cell multi-omics using generative {AI},''
  \emph{bioRxiv}, pp. 2023--04, 2023.

\bibitem{du2023diffusion}
H.~Du, Z.~Li, D.~Niyato, J.~Kang, Z.~Xiong, H.~Huang, and S.~Mao,
  ``Diffusion-based reinforcement learning for edge-enabled {AI}-generated
  content services,'' \emph{arXiv preprint arXiv:2303.13052}, 2023.

\bibitem{314}
G.~Tosato, C.~M. Dalbagno, and F.~Fumagalli, ``{EEG} synthetic data generation
  using probabilistic diffusion models,'' \emph{arXiv preprint
  arXiv:2303.06068}, 2023.

\bibitem{296}
K.~Gong, K.~Johnson, G.~El~Fakhri \emph{et~al.}, ``{PET} image denoising based
  on denoising diffusion probabilistic model,'' \emph{Eur. J. Nucl. Med. Mol.
  Imaging}, pp. 1--11, 2023.

\bibitem{432}
T.~Hu, F.~Hong, and Z.~Liu, ``{StructLDM}: Structured latent diffusion for {3D}
  human generation,'' \emph{arXiv preprint arXiv:2404.01241}, 2024.

\bibitem{294}
X.~Meng, Y.~Gu, Y.~Pan \emph{et~al.}, ``A novel unified conditional score-based
  generative framework for multi-modal medical image completion,'' \emph{arXiv
  preprint arXiv:2207.03430}, 2022.

\bibitem{321}
M.~Hajij, G.~Zamzmi, R.~Paul \emph{et~al.}, ``Normalizing flow for synthetic
  medical images generation,'' in \emph{Proc. HI-POCT}, 2022, pp. 46--49.

\bibitem{du2023semantic}
H.~Du, J.~Wang, D.~Niyato, J.~Kang, Z.~Xiong, J.~Zhang, and X.~Shen, ``Semantic
  communications for wireless sensing: {RIS}-aided encoding and self-supervised
  decoding,'' \emph{IEEE J. Sel. Areas Commun.}, to appear, 2023.

\bibitem{412}
J.~Li, C.~Yi, J.~Chen \emph{et~al.}, ``Joint trajectory planning, application
  placement, and energy renewal for {UAV}-assisted {MEC}: A
  triple-learner-based approach,'' \emph{IEEE Internet Things J.}, vol.~10,
  no.~15, pp. 13\,622--13\,636, 2023.

\bibitem{413}
J.~Chen, C.~Yi, R.~Wang \emph{et~al.}, ``Learning aided joint sensor activation
  and mobile charging vehicle scheduling for energy-efficient wrsn-based
  industrial {IoT},'' \emph{IEEE Trans. Veh. Technol.}, vol.~72, no.~4, pp.
  5064--5078, 2023.

\bibitem{408}
Y.~Shi, C.~Yi, B.~Chen \emph{et~al.}, ``Closed-loop control of edge-cloud
  collaboration enabled {IIoT}: An online optimization approach,'' in
  \emph{Proc. IEEE ICC}.\hskip 1em plus 0.5em minus 0.4em\relax IEEE, 2022, pp.
  5682--5687.

\bibitem{246}
H.~Du, J.~Wang, D.~Niyato \emph{et~al.}, ``{AI}-generated incentive mechanism
  and full-duplex semantic communications for information sharing,'' \emph{IEEE
  J. Sel. Areas in Commun.}, vol.~41, no.~9, pp. 2981--2997, 2023.

\bibitem{385}
C.~Liang, H.~Du, Y.~Sun \emph{et~al.}, ``Generative {AI}-driven semantic
  communication networks: Architecture, technologies and applications,''
  \emph{arXiv preprint arXiv:2401.00124}, 2023.

\bibitem{386}
H.~Du, G.~Liu, D.~Niyato \emph{et~al.}, ``Generative {AI}-aided joint
  training-free secure semantic communications via multi-modal prompts,''
  \emph{arXiv preprint arXiv:2309.02616}, 2023.

\bibitem{332}
X.~Wei, D.~Wu, L.~Zhou \emph{et~al.}, ``Cross-modal communication technology: A
  survey,'' \emph{Fundam. Res.}, 2023.

\bibitem{39}
J.~Corral-Acero, F.~Margara, M.~Marciniak \emph{et~al.}, ``The ‘digital
  twin’ to enable the vision of precision cardiology,'' \emph{Eur. Heart J.},
  vol.~41, no.~48, pp. 4556--4564, 2020.

\bibitem{316}
D.~Hazra and Y.-C. Byun, ``{SynSigGAN}: Generative adversarial networks for
  synthetic biomedical signal generation,'' \emph{Biology}, vol.~9, no.~12, p.
  441, 2020.

\bibitem{367}
M.~A. Pimentel, A.~E. Johnson, P.~H. Charlton \emph{et~al.}, ``Toward a robust
  estimation of respiratory rate from pulse oximeters,'' \emph{IEEE Trans.
  Biomed. Eng.}, vol.~64, no.~8, pp. 1914--1923, 2016.

\bibitem{288}
J.~M.~L. Alcaraz and N.~Strodthoff, ``Diffusion-based conditional {ECG}
  generation with structured state space models,'' \emph{Comput. Biol. Med.},
  p. 107115, 2023.

\bibitem{318}
------, ``Diffusion-based time series imputation and forecasting with
  structured state space models,'' \emph{Trans. Mach. Learn. Res.}, 2022.

\bibitem{317}
P.~Wagner, N.~Strodthoff, R.-D. Bousseljot \emph{et~al.}, ``{PTB-XL}, a large
  publicly available electrocardiography dataset,'' \emph{Sci. data}, vol.~7,
  no.~1, p. 154, 2020.

\bibitem{319}
P.~A. Moghadam, S.~Van~Dalen, K.~C. Martin \emph{et~al.}, ``A morphology
  focused diffusion probabilistic model for synthesis of histopathology
  images,'' in \emph{Proc. IEEE/CVF CVPR}, 2023, pp. 2000--2009.

\bibitem{368}
A.~B. Levine, J.~Peng, D.~Farnell \emph{et~al.}, ``Synthesis of diagnostic
  quality cancer pathology images by generative adversarial networks,''
  \emph{J. pathol.}, vol. 252, no.~2, pp. 178--188, 2020.

\bibitem{322}
L.~Dinh, J.~Sohl-Dickstein, and S.~Bengio, ``Density estimation using real
  {NVP},'' \emph{arXiv preprint arXiv:1605.08803}, 2016.

\bibitem{323}
R.~Summers, ``Nih chest x-ray dataset of 14 common thorax disease categories,''
  \emph{NIH Clinical Center: Bethesda, MD, USA}, 2019.

\bibitem{324}
P.~Tschandl, C.~Rosendahl, and H.~Kittler, ``The {HAM10000} dataset, a large
  collection of multi-source dermatoscopic images of common pigmented skin
  lesions,'' \emph{Sci. data}, vol.~5, no.~1, pp. 1--9, 2018.

\bibitem{326}
Y.~Du, R.~Kips, A.~Pumarola \emph{et~al.}, ``Avatars grow legs: Generating
  smooth human motion from sparse tracking inputs with diffusion model,'' in
  \emph{Proc. IEEE/CVF CVPR}, 2023, pp. 481--490.

\bibitem{394}
S.~Okegbile, J.~Cai, J.~Wu \emph{et~al.}, ``A prediction-enhanced
  physical-to-virtual twin connectivity framework for human digital twin,''
  \emph{Authorea Preprints}, 2023.

\bibitem{422}
K.~Peng, P.~Xiao, S.~Wang \emph{et~al.}, ``{SCOF}: Security-aware computation
  offloading using federated reinforcement learning in industrial internet of
  things with edge computing,'' \emph{IEEE Trans. Serv. Comput.}, pp. 1--13,
  2024.

\bibitem{423}
------, ``{AoI}-aware partial computation offloading in {IIoT} with edge
  computing: A deep reinforcement learning based approach,'' \emph{IEEE Trans.
  Cloud Comput.}, vol.~11, no.~4, pp. 3766--3777, 2023.

\bibitem{402}
W.~Zhu, R.~Chen, C.~Yi \emph{et~al.}, ``Edge-assisted video transmission with
  adaptive key frame selection: A hierarchical {DRL} approach,'' in \emph{Proc.
  BSC}.\hskip 1em plus 0.5em minus 0.4em\relax IEEE, 2023, pp. 89--94.

\bibitem{yang2022semantic}
W.~Yang, H.~Du, Z.~Q. Liew, W.~Y.~B. Lim, Z.~Xiong, D.~Niyato, X.~Chi, X.~S.
  Shen, and C.~Miao, ``Semantic communications for future {I}nternet:
  {F}undamentals, applications, and challenges,'' \emph{IEEE Communications
  Surveys \& Tutorials}, 2023.

\bibitem{du2023user}
H.~Du, R.~Zhang, D.~Niyato, J.~Kang, Z.~Xiong, S.~Cui, X.~Shen, and D.~I. Kim,
  ``User-centric interactive {AI} for distributed diffusion model-based
  {AI}-generated content,'' \emph{arXiv preprint arXiv:2311.11094}, 2023.

\bibitem{328}
A.~D. Raha, M.~S. Munir, A.~Adhikary \emph{et~al.}, ``Generative {AI}-driven
  semantic communication framework for {NextG} wireless network,'' \emph{arXiv
  preprint arXiv:2310.09021}, 2023.

\bibitem{329}
C.~Zhang, D.~Han, Y.~Qiao \emph{et~al.}, ``Faster segment anything: Towards
  lightweight {SAM} for mobile applications,'' \emph{arXiv preprint
  arXiv:2306.14289}, 2023.

\bibitem{331}
E.~Grassucci, S.~Barbarossa, and D.~Comminiello, ``Generative semantic
  communication: Diffusion models beyond bit recovery,'' \emph{arXiv preprint
  arXiv:2306.04321}, 2023.

\bibitem{334}
H.~Liu, D.~Guo, X.~Zhang \emph{et~al.}, ``Toward image-to-tactile cross-modal
  perception for visually impaired people,'' \emph{IEEE Trans. Autom. Sci.
  Eng.}, vol.~18, no.~2, pp. 521--529, 2021.

\bibitem{335}
T.~Kim, M.~Cha, H.~Kim \emph{et~al.}, ``Learning to discover cross-domain
  relations with generative adversarial networks,'' in \emph{Proc. ICML}.\hskip
  1em plus 0.5em minus 0.4em\relax PMLR, 2017, pp. 1857--1865.

\bibitem{336}
Y.~Fang, X.~Zhang, W.~Xu \emph{et~al.}, ``Bidirectional visual-tactile
  cross-modal generation using latent feature space flow model,''
  \emph{Available at SSRN 4593113}.

\bibitem{337}
A.~Li, Y.~Chen, S.~Ni \emph{et~al.}, ``Haptic signal reconstruction in ehealth
  internet of things,'' \emph{IEEE Internet Things J.}, vol.~9, no.~18, pp.
  17\,047--17\,057, 2021.

\bibitem{40}
B.~R. Barricelli, E.~Casiraghi, J.~Gliozzo \emph{et~al.}, ``Human digital twin
  for fitness management,'' \emph{IEEE Access}, vol.~8, pp. 26\,637--26\,664,
  2020.

\bibitem{358}
S.~Phung, A.~Kumar, and J.~Kim, ``A deep learning technique for imputing
  missing healthcare data,'' in \emph{Proc. IEEE EMBC}.\hskip 1em plus 0.5em
  minus 0.4em\relax IEEE, 2019, pp. 6513--6516.

\bibitem{84}
W.~Dong, D.~Y.~T. Fong, J.-s. Yoon \emph{et~al.}, ``Generative adversarial
  networks for imputing missing data for big data clinical research,''
  \emph{BMC medical res. methodol.}, vol.~21, pp. 1--10, 2021.

\bibitem{365}
P.~Hayati~Rezvan, K.~J. Lee, and J.~A. Simpson, ``The rise of multiple
  imputation: a review of the reporting and implementation of the method in
  medical research,'' \emph{BMC med. res. methodol.}, vol.~15, pp. 1--14, 2015.

\bibitem{366}
D.~J. Stekhoven and P.~B{\"u}hlmann, ``{MissForest}—non-parametric missing
  value imputation for mixed-type data,'' \emph{Bioinformatics}, vol.~28,
  no.~1, pp. 112--118, 2012.

\bibitem{291}
Q.~Lyu and G.~Wang, ``Conversion between {CT} and mri images using diffusion
  and score-matching models,'' \emph{arXiv preprint arXiv:2209.12104}, 2022.

\bibitem{292}
C.~Saharia, J.~Ho, W.~Chan \emph{et~al.}, ``Image super-resolution via
  iterative refinement,'' \emph{IEEE Trans. Pattern Anal. Mach. Intell.},
  vol.~45, no.~4, pp. 4713--4726, 2022.

\bibitem{293}
Y.~Song, J.~Sohl-Dickstein, D.~P. Kingma \emph{et~al.}, ``Score-based
  generative modeling through stochastic differential equations,'' \emph{Proc.
  ICLR}, 2021.

\bibitem{369}
------, ``Score-based generative modeling through stochastic differential
  equations,'' \emph{arXiv preprint arXiv:2011.13456}, 2020.

\bibitem{370}
T.~Nyholm, S.~Svensson, S.~Andersson \emph{et~al.}, ``{MR and CT} data with
  multiobserver delineations of organs in the pelvic area—part of the gold
  atlas project,'' \emph{Med. phys.}, vol.~45, no.~3, pp. 1295--1300, 2018.

\bibitem{371}
I.~Gulrajani, F.~Ahmed, M.~Arjovsky \emph{et~al.}, ``Improved training of
  wasserstein {GANs},'' \emph{Proc. NeurIPS}, vol.~30, 2017.

\bibitem{372}
P.~Huang, D.~Li, Z.~Jiao \emph{et~al.}, ``{CoCa-GAN}:
  common-feature-learning-based context-aware generative adversarial network
  for glioma grading,'' in \emph{Proc. MICCAI}.\hskip 1em plus 0.5em minus
  0.4em\relax Springer, 2019, pp. 155--163.

\bibitem{373}
T.~Zhou, H.~Fu, G.~Chen \emph{et~al.}, ``{Hi-Net}: hybrid-fusion network for
  multi-modal mr image synthesis,'' \emph{IEEE trans. med. imaging}, vol.~39,
  no.~9, pp. 2772--2781, 2020.

\bibitem{374}
A.~Sharma and G.~Hamarneh, ``Missing {MRI} pulse sequence synthesis using
  multi-modal generative adversarial network,'' \emph{IEEE trans. med.
  imaging}, vol.~39, no.~4, pp. 1170--1183, 2019.

\bibitem{375}
X.~Liu, F.~Xing, G.~El~Fakhri \emph{et~al.}, ``A unified conditional
  disentanglement framework for multimodal brain mr image translation,'' in
  \emph{Proc. ISBI}.\hskip 1em plus 0.5em minus 0.4em\relax IEEE, 2021, pp.
  10--14.

\bibitem{376}
A.~Chartsias, T.~Joyce, M.~V. Giuffrida \emph{et~al.}, ``Multimodal {MR}
  synthesis via modality-invariant latent representation,'' \emph{IEEE trans.
  med. imaging}, vol.~37, no.~3, pp. 803--814, 2017.

\bibitem{295}
S.~Kazeminia, C.~Baur, A.~Kuijper \emph{et~al.}, ``{GANs} for medical image
  analysis,'' \emph{Artif. Intell. Med.}, vol. 109, p. 101938, 2020.

\bibitem{359}
H.-T. Chiang, Y.-Y. Hsieh, S.-W. Fu \emph{et~al.}, ``Noise reduction in {ECG}
  signals using fully convolutional denoising autoencoders,'' \emph{IEEE
  Access}, vol.~7, pp. 60\,806--60\,813, 2019.

\bibitem{360}
S.~Nasrin, M.~Z. Alom, R.~Burada \emph{et~al.}, ``Medical image denoising with
  recurrent residual {U-Net (R2U-Net)} base auto-encoder,'' in \emph{Proc. IEEE
  NAECON}.\hskip 1em plus 0.5em minus 0.4em\relax IEEE, 2019, pp. 345--350.

\bibitem{377}
P.~Isola, J.-Y. Zhu, T.~Zhou \emph{et~al.}, ``Image-to-image translation with
  conditional adversarial networks,'' in \emph{Proc. CVPR}, 2017, pp.
  1125--1134.

\bibitem{297}
T.~Xiang, M.~Yurt, A.~B. Syed \emph{et~al.}, ``{DDM$^2$} : Self-supervised
  diffusion mri denoising with generative diffusion models,'' \emph{Proc.
  ICLR}, 2023.

\bibitem{434}
K.~Kim and J.~C. Ye, ``{Noise2Score}: tweedie’s approach to self-supervised
  image denoising without clean images,'' \emph{Proc. NeurIPS}, vol.~34, pp.
  864--874, 2021.

\bibitem{361}
A.~Majeed, ``Attribute-centric anonymization scheme for improving user privacy
  and utility of publishing e-health data,'' \emph{J. King Saud Univ. - Comput.
  Inf. Sci.}, vol.~31, no.~4, pp. 426--435, 2019.

\bibitem{362}
A.~Aminifar, F.~Rabbi, V.~K.~I. Pun \emph{et~al.}, ``Diversity-aware
  anonymization for structured health data,'' in \emph{Proc. IEEE EMBC}.\hskip
  1em plus 0.5em minus 0.4em\relax IEEE, 2021, pp. 2148--2154.

\bibitem{298}
J.~Yoon, L.~N. Drumright, and M.~van~der Schaar, ``Anonymization through data
  synthesis using generative adversarial networks {(ADS-GAN)},'' \emph{IEEE J.
  Biomed. Health Inform.}, vol.~24, no.~8, pp. 2378--2388, 2020.

\bibitem{380}
J.~Jordon, J.~Yoon, and M.~Van Der~Schaar, ``{PATE-GAN}: Generating synthetic
  data with differential privacy guarantees,'' in \emph{Proc. ICLR}, 2018.

\bibitem{269}
M.~Bordukova, N.~Makarov, R.~Rodriguez-Esteban \emph{et~al.}, ``Generative
  artificial intelligence empowers digital twins in drug discovery and clinical
  trials,'' \emph{Expert Opin. Drug Discov.}, pp. 1--10, 2023.

\bibitem{65}
B.~S. Center, ``Alya {Red}: A computational heart,''
  \url{https://www.bsc.es/news/bsc-news/alya-red-computational-heart}, 2023.

\bibitem{251}
M.~Lotfollahi, F.~A. Wolf, and F.~J. Theis, ``{scGen} predicts single-cell
  perturbation responses,'' \emph{Nat. Methods}, vol.~16, no.~8, pp. 715--721,
  2019.

\bibitem{252}
M.~Lotfollahi, A.~Klimovskaia~Susmelj, C.~De~Donno \emph{et~al.}, ``Predicting
  cellular responses to complex perturbations in high-throughput screens,''
  \emph{Mol. Syst. Biol.}, p. e11517, 2023.

\bibitem{279}
R.~M. Donovan-Maiye, J.~M. Brown, C.~K. Chan \emph{et~al.}, ``A deep generative
  model of {3D} single-cell organization,'' \emph{PLOS Comput. Biol.}, vol.~18,
  no.~1, p. e1009155, 2022.

\bibitem{280}
L.~Jose, S.~Liu, C.~Russo \emph{et~al.}, ``Generative adversarial networks in
  digital pathology and histopathological image processing: A review,''
  \emph{J. Pathol. Inform.}, vol.~12, no.~1, p.~43, 2021.

\bibitem{281}
A.~C. Quiros, R.~Murray-Smith, and K.~Yuan, ``{PathologyGAN}: Learning deep
  representations of cancer tissue,'' in \emph{Proc. PMLR}, vol. 121.\hskip 1em
  plus 0.5em minus 0.4em\relax PMLR, 06--08 Jul 2020, pp. 669--695.

\bibitem{282}
A.~Brock, J.~Donahue, and K.~Simonyan, ``Large scale {GAN} training for high
  fidelity natural image synthesis,'' \emph{arXiv preprint arXiv:1809.11096},
  2018.

\bibitem{283}
T.~Karras, S.~Laine, and T.~Aila, ``A style-based generator architecture for
  generative adversarial networks,'' in \emph{Proc. IEEE/CVF CVPR}, 2019, pp.
  4401--4410.

\bibitem{284}
A.~Jolicoeur-Martineau, ``The relativistic discriminator: a key element missing
  from standard gan,'' \emph{arXiv preprint arXiv:1807.00734}, 2018.

\bibitem{256}
H.~Ahmadian, P.~Mageswaran, B.~A. Walter \emph{et~al.}, ``Toward an artificial
  intelligence-assisted framework for reconstructing the digital twin of
  vertebra and predicting its fracture response,'' \emph{Int. J. Numer. Method
  Biomed. Eng.}, vol.~38, no.~6, p. e3601, 2022.

\bibitem{381}
A.~Radford, L.~Metz, and S.~Chintala, ``Unsupervised representation learning
  with deep convolutional generative adversarial networks,'' \emph{arXiv
  preprint arXiv:1511.06434}, 2015.

\bibitem{254}
X.~Xing, J.~Del~Ser, Y.~Wu \emph{et~al.}, ``{HDL}: Hybrid deep learning for the
  synthesis of myocardial velocity maps in digital twins for cardiac
  analysis,'' \emph{IEEE J. Biomed. Health. Inform.}, 2022.

\bibitem{364}
R.~Martinez-Velazquez, R.~Gamez, and A.~El~Saddik, ``Cardio twin: A digital
  twin of the human heart running on the edge,'' in \emph{Proc. IEEE
  MeMeA}.\hskip 1em plus 0.5em minus 0.4em\relax IEEE, 2019, pp. 1--6.

\bibitem{363}
X.~Li, Z.~Gong, H.~Yin \emph{et~al.}, ``A {3D} deep supervised densely network
  for small organs of human temporal bone segmentation in {CT} images,''
  \emph{Neural Netw.}, vol. 124, pp. 75--85, 2020.

\bibitem{301}
N.~J. Dhinagar, S.~I. Thomopoulos, E.~Laltoo \emph{et~al.}, ``Efficiently
  training vision transformers on structural {MRI} scans for alzheimer's
  disease detection,'' \emph{arXiv preprint arXiv:2303.08216}, 2023.

\bibitem{302}
Y.~Dong, M.~Zhang, L.~Qiu \emph{et~al.}, ``An arrhythmia classification model
  based on vision transformer with deformable attention,''
  \emph{Micromachines}, vol.~14, no.~6, p. 1155, 2023.

\bibitem{382}
C.~Che, P.~Zhang, M.~Zhu \emph{et~al.}, ``Constrained transformer network for
  {ECG} signal processing and arrhythmia classification,'' \emph{BMC Med.
  Inform. Decis. Mak.}, vol.~21, no.~1, pp. 1--13, 2021.

\bibitem{304}
A.~Rahman, J.~M.~J. Valanarasu, I.~Hacihaliloglu \emph{et~al.}, ``Ambiguous
  medical image segmentation using diffusion models,'' in \emph{Proc. IEEE/CVF
  CVPR}, 2023, pp. 11\,536--11\,546.

\bibitem{303}
B.~Kim, Y.~Oh, and J.~C. Ye, ``Diffusion adversarial representation learning
  for self-supervised vessel segmentation,'' \emph{Proc. ICLR}, 2023.

\bibitem{305}
X.~Guo, J.~W. Gichoya, S.~Purkayastha \emph{et~al.}, ``{CVAD}: An anomaly
  detector for medical images based on cascade {VAE},'' in \emph{Proc.
  MILLanD}.\hskip 1em plus 0.5em minus 0.4em\relax Springer, 2022, pp.
  187--196.

\bibitem{306}
T.~Nakao, S.~Hanaoka, Y.~Nomura \emph{et~al.}, ``Unsupervised deep anomaly
  detection in chest radiographs,'' \emph{J. Digit. Imaging}, vol.~34, pp.
  418--427, 2021.

\bibitem{300}
J.~Wolleb, F.~Bieder, R.~Sandk{\"u}hler \emph{et~al.}, ``Diffusion models for
  medical anomaly detection,'' in \emph{Proc. MICCAI}.\hskip 1em plus 0.5em
  minus 0.4em\relax Springer, 2022, pp. 35--45.

\bibitem{383}
J.~Irvin, P.~Rajpurkar, M.~Ko \emph{et~al.}, ``Chexpert: A large chest
  radiograph dataset with uncertainty labels and expert comparison,'' in
  \emph{Proc. AAAI}, vol.~33, no.~01, 2019, pp. 590--597.

\bibitem{384}
A.~Johnson, L.~Bulgarelli, T.~Pollard \emph{et~al.}, ``{MIMIC-IV},''
  \emph{PhysioNet}, 2021.

\bibitem{307}
W.~Hu and S.~Y. Wang, ``Predicting glaucoma progression requiring surgery using
  clinical free-text notes and transfer learning with transformers,''
  \emph{Transl. Vis. Sci. Techn.}, vol.~11, no.~3, pp. 37--37, 2022.

\bibitem{308}
J.~Devlin, M.-W. Chang, K.~Lee \emph{et~al.}, ``Bert: Pre-training of deep
  bidirectional transformers for language understanding,'' \emph{arXiv preprint
  arXiv:1810.04805}, 2018.

\bibitem{309}
J.~Lee, W.~Yoon, S.~Kim \emph{et~al.}, ``{BioBERT}: a pre-trained biomedical
  language representation model for biomedical text mining,''
  \emph{Bioinformatics}, vol.~36, no.~4, pp. 1234--1240, 2020.

\bibitem{310}
Y.~Liu, M.~Ott, N.~Goyal, J.~Du, M.~Joshi, D.~Chen, O.~Levy, M.~Lewis,
  L.~Zettlemoyer, and V.~Stoyanov, ``{RoBERTa}: A robustly optimized bert
  pretraining approach,'' \emph{arXiv preprint arXiv:1907.11692}, 2019.

\bibitem{311}
V.~Sanh, L.~Debut, J.~Chaumond \emph{et~al.}, ``{DistilBERT}, a distilled
  version of {BERT}: smaller, faster, cheaper and lighter,'' \emph{arXiv
  preprint arXiv:1910.01108}, 2019.

\bibitem{409}
Z.~Tu, K.~Zhu \emph{et~al.}, ``Blockchain-based privacy-preserving dynamic
  spectrum sharing,'' in \emph{Proc. WASA}.\hskip 1em plus 0.5em minus
  0.4em\relax Springer, 2020, pp. 444--456.

\bibitem{393}
S.~Okegbile, O.~Talabi, H.~Gao \emph{et~al.}, ``{FLeS}: A federated
  learning-enhanced semantic communication framework for mobile {AIGC}-driven
  human digital twins,'' \emph{Authorea Preprints}, 2023.

\bibitem{66}
S.~D. Okegbile, J.~Cai, H.~Zheng \emph{et~al.}, ``Differentially private
  federated multi-task learning framework for enhancing human-to-virtual
  connectivity in human digital twin,'' \emph{IEEE J. Sel. Areas Commun.}, pp.
  1--1, 2023.

\bibitem{407}
Y.~Shi, C.~Yi, B.~Chen \emph{et~al.}, ``A two-timescale online optimization for
  balancing service migration and task rerouting in {MEC},'' in \emph{IEEE
  Proc. GLOBECOM}.\hskip 1em plus 0.5em minus 0.4em\relax IEEE, 2023, pp.
  5500--5505.

\bibitem{399}
R.~Chen, C.~Yi, K.~Zhu \emph{et~al.}, ``A three-party hierarchical game for
  physical layer security aware wireless communications with dynamic trilateral
  coalitions,'' \emph{IEEE Trans. Wirel. Commun.}, 2023.

\bibitem{400}
------, ``A {DRL}-based hierarchical game for physical layer security with
  dynamic trilateral coalitions,'' in \emph{Proc. IEEE ICC}.\hskip 1em plus
  0.5em minus 0.4em\relax IEEE, 2023, pp. 4495--4500.

\bibitem{401}
H.~Zhou, R.~Chen, C.~Yi \emph{et~al.}, ``A three-party repeated coalition
  formation game for {PLS} in wireless communications with {IRSs},''
  \emph{arXiv preprint arXiv:2402.11500}, 2024.

\bibitem{341}
Z.~Wang, S.~Stavrakis, and B.~Yao, ``Hierarchical deep learning with generative
  adversarial network for automatic cardiac diagnosis from {ECG} signals,''
  \emph{Comput. Biol. Med.}, vol. 155, p. 106641, 2023.

\bibitem{newstart_425}
B.~Zhou, S.~Liu, B.~Hooi \emph{et~al.}, ``{BeatGAN}: Anomalous rhythm detection
  using adversarially generated time series,'' in \emph{Proc. IJCAI}, vol.
  2019, 2019, pp. 4433--4439.

\bibitem{391}
Paige, ``Paige,'' \url{https://paige.ai}, 2024.

\bibitem{390}
P.~Raciti, J.~Sue, J.~A. Retamero \emph{et~al.}, ``Clinical validation of
  artificial intelligence--augmented pathology diagnosis demonstrates
  significant gains in diagnostic accuracy in prostate cancer detection,''
  \emph{Arch. of Pathol. Lab. Med}, vol. 147, no.~10, pp. 1178--1185, 2023.

\bibitem{389}
L.~M. da~Silva, E.~M. Pereira, P.~G. Salles \emph{et~al.}, ``Independent
  real-world application of a clinical-grade automated prostate cancer
  detection system,'' \emph{J. Pathol.}, vol. 254, no.~2, pp. 147--158, 2021.

\bibitem{346}
T.~N. Jarada, J.~G. Rokne, and R.~Alhajj, ``{SNF--CVAE}: computational method
  to predict drug--disease interactions using similarity network fusion and
  collective variational autoencoder,'' \emph{Knowl. Based Syst.}, vol. 212, p.
  106585, 2021.

\bibitem{427}
J.~Davis and M.~Goadrich, ``The relationship between precision-recall and {ROC}
  curves,'' in \emph{Proc. ICML}, 2006, pp. 233--240.

\bibitem{428}
M.~G. Ozsoy, T.~{\"O}zyer, F.~Polat \emph{et~al.}, ``Realizing drug
  repositioning by adapting a recommendation system to handle the process,''
  \emph{BMC bioinf.}, vol.~19, pp. 1--14, 2018.

\bibitem{239}
I.~Medicine, ``Pharma.ai,'' \url{ https://insilico.com }, 2024.

\bibitem{388}
F.~Ren, A.~Aliper, J.~Chen \emph{et~al.}, ``A small-molecule {TNIK} inhibitor
  targets fibrosis in preclinical and clinical models,'' \emph{Nat.
  Biotechnol.}, pp. 1--13, 2024.

\bibitem{338}
C.~Mennella, U.~Maniscalco, G.~De~Pietro \emph{et~al.}, ``Generating a novel
  synthetic dataset for rehabilitation exercises using pose-guided conditioned
  diffusion models: A quantitative and qualitative evaluation,'' \emph{Comput.
  Biol. Med.}, vol. 167, p. 107665, 2023.

\bibitem{339}
L.~Zhang, A.~Rao, and M.~Agrawala, ``Adding conditional control to
  text-to-image diffusion models,'' in \emph{Proc. IEEE/CVF ICCV}, 2023, pp.
  3836--3847.

\bibitem{185}
X.~Du, Y.~Liu, Z.~Lu \emph{et~al.}, ``A low-latency communication design for
  brain simulations,'' \emph{IEEE Netw.}, vol.~36, no.~2, pp. 8--15, 2022.

\bibitem{186}
W.~Lu \emph{et~al.}, ``The human digital twin brain in the resting state and in
  action,'' \emph{arXiv preprint arXiv:2211.15963}, 2022.

\end{thebibliography}
\begin{IEEEbiography}[{\includegraphics[width=1in,height=1.25in,clip,keepaspectratio]{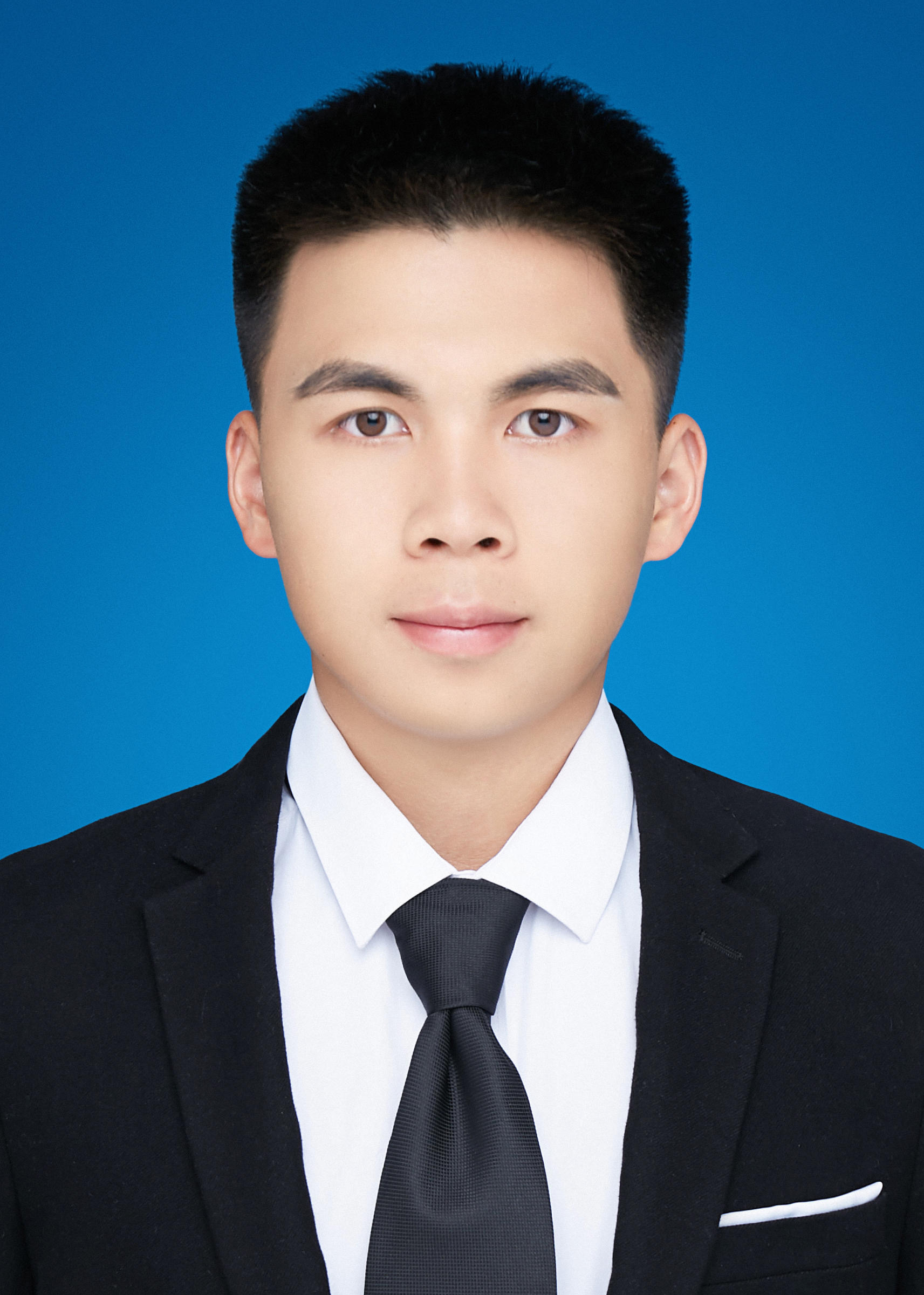}}]{Jiayuan Chen} received the M.S. degree with the College of Computer Science and Technology, Nanjing University of Aeronautics and Astronautics (NUAA), Nanjing, China, where he is currently pursuing the Ph.D. degree in Computer Science and Technology. His research interests include reinforcement learning, generative AI and network management.
\end{IEEEbiography}

\begin{IEEEbiography}[{\includegraphics[width=1in,height=1.25in,clip,keepaspectratio]{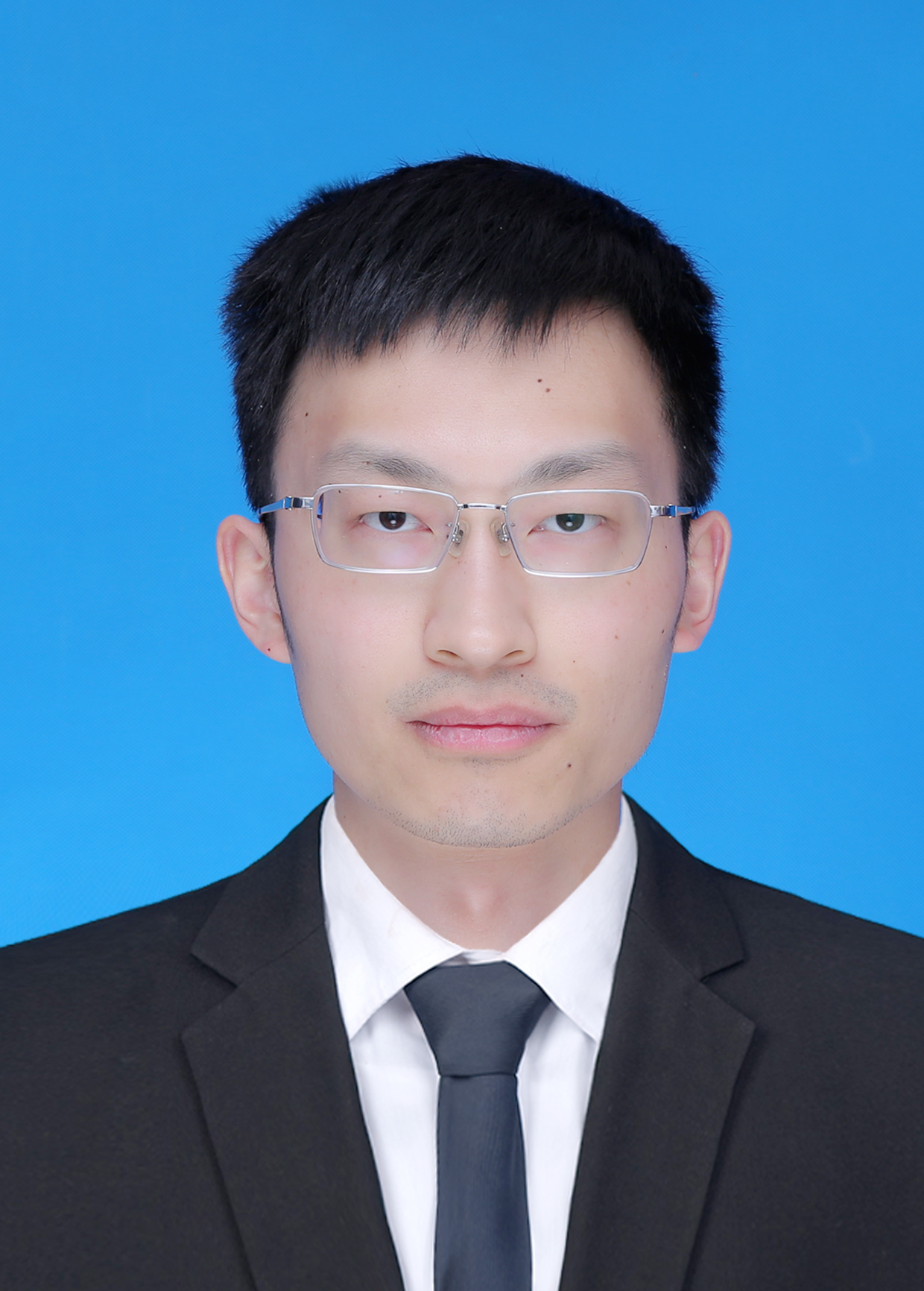}}]{You Shi} received the M.S. degree from the School of Computer Science and Communication Engineering, Jiangsu University, Zhenjiang, China, in 2020. He is pursuing a Ph.D. at the College of Computer Science and Technology, Nanjing University of Aeronautics and Astronautics (NUAA), Nanjing, China. His main research interests include mobile edge computing, online optimization, and service deployment.
\end{IEEEbiography}

\begin{IEEEbiography}[{\includegraphics[width=1in,height=1.25in,clip,keepaspectratio]{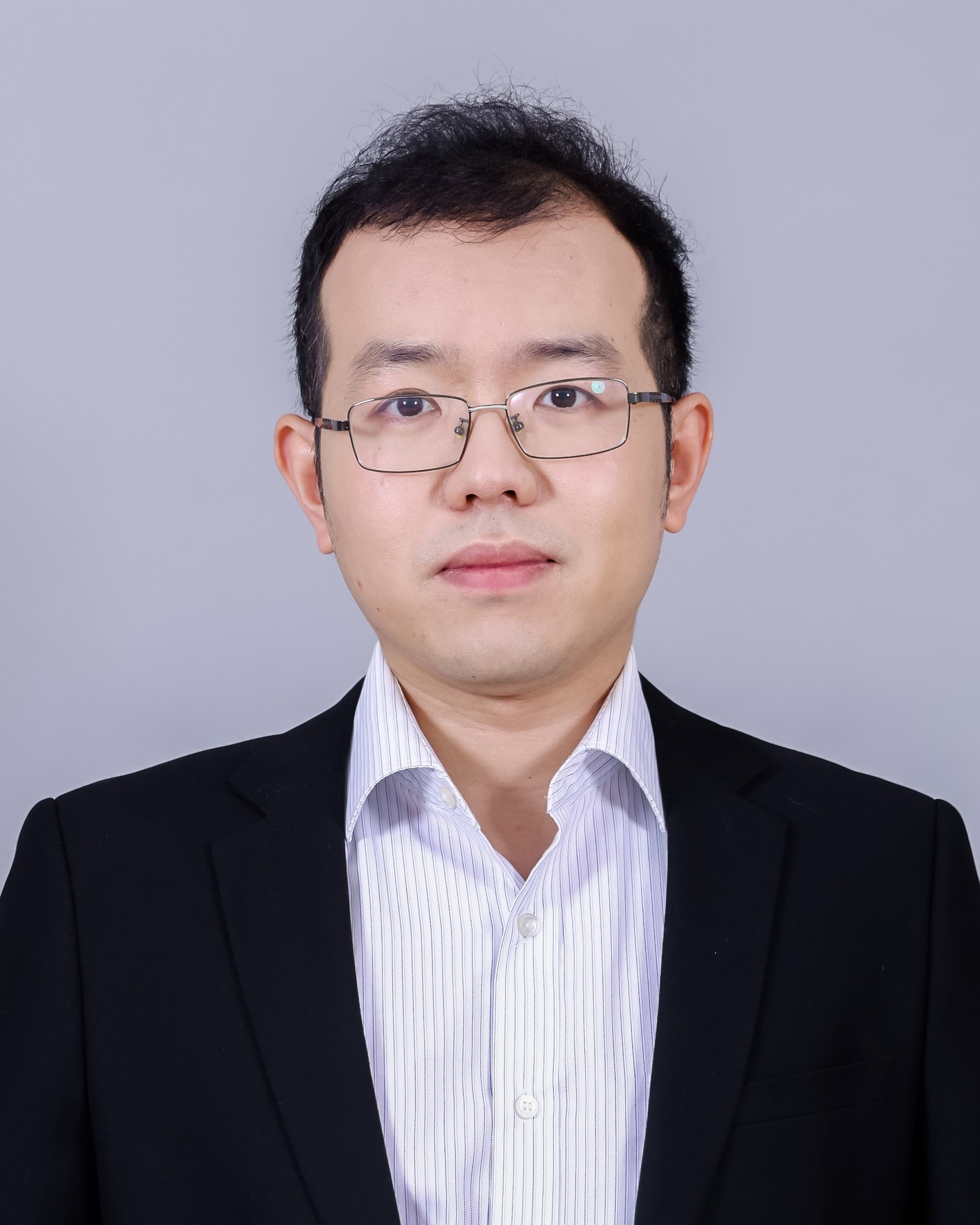}}]{Changyan Yi} (Member, IEEE) received the Ph.D. degree from the Department of Electrical and Computer Engineering, University of Manitoba, MB, Canada, in 2018. He is currently a Professor with the College of Computer Science and Technology, Nanjing University of Aeronautics and Astronautics (NUAA), Nanjing, China. His research interests include stochastic optimization, game theory, incentive mechanism, with applications in various communication and networking systems.
\end{IEEEbiography}

\begin{IEEEbiography}[{\includegraphics[width=1in,height=1.25in, clip,keepaspectratio]{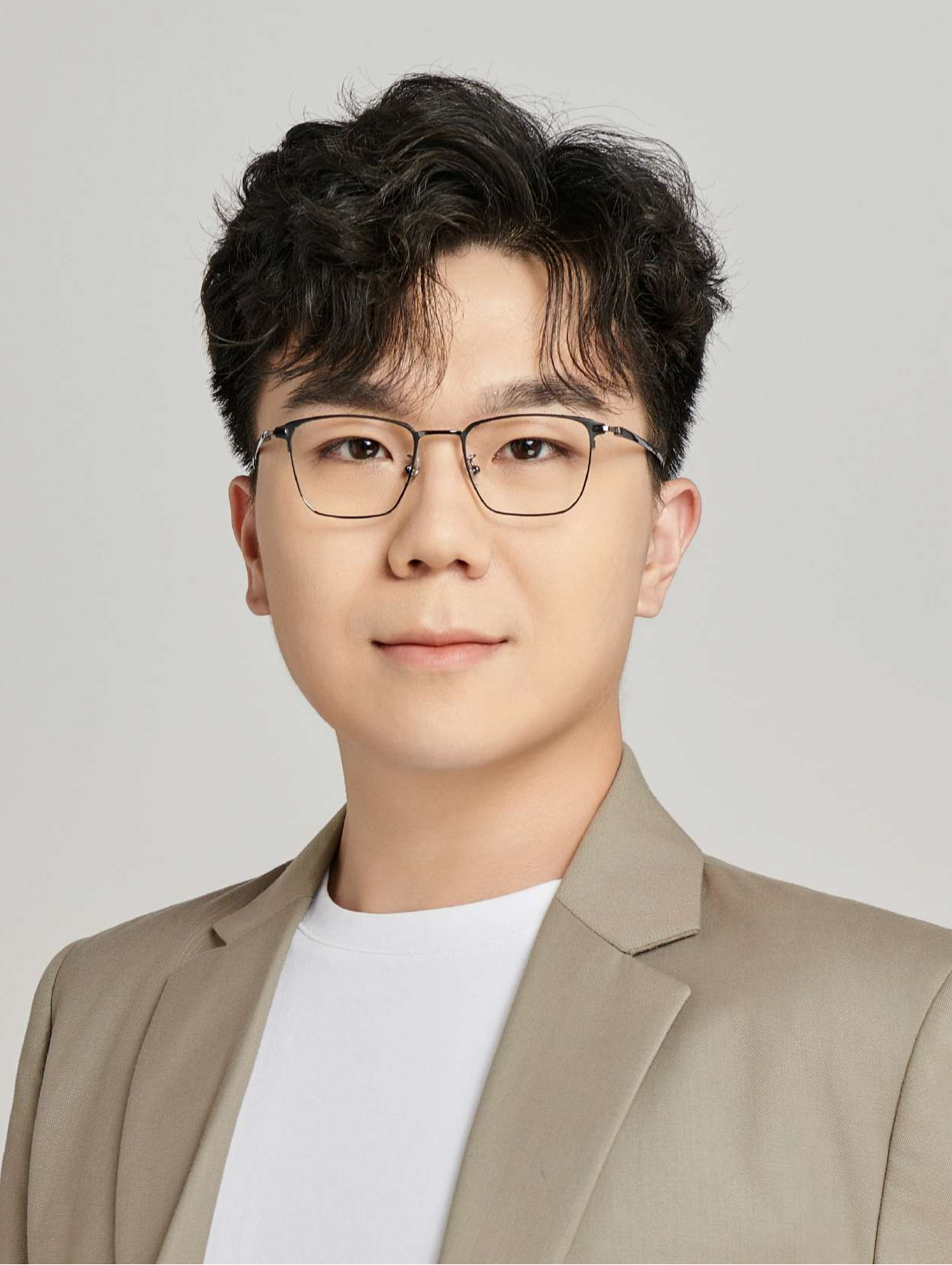}}]{Hongyang Du}
	received the BEng degree from the School of Electronic and Information Engineering, Beijing Jiaotong University, Beijing, in 2021, and the PhD degree from the Interdisciplinary Graduate Program at the College of Computing and Data Science, Energy Research Institute @ NTU, Nanyang Technological University, Singapore, in 2024. His research interests include edge intelligence, generative AI, and semantic communications.
\end{IEEEbiography}

\begin{IEEEbiography}[{\includegraphics[width=1in,height=1.25in,clip,keepaspectratio]{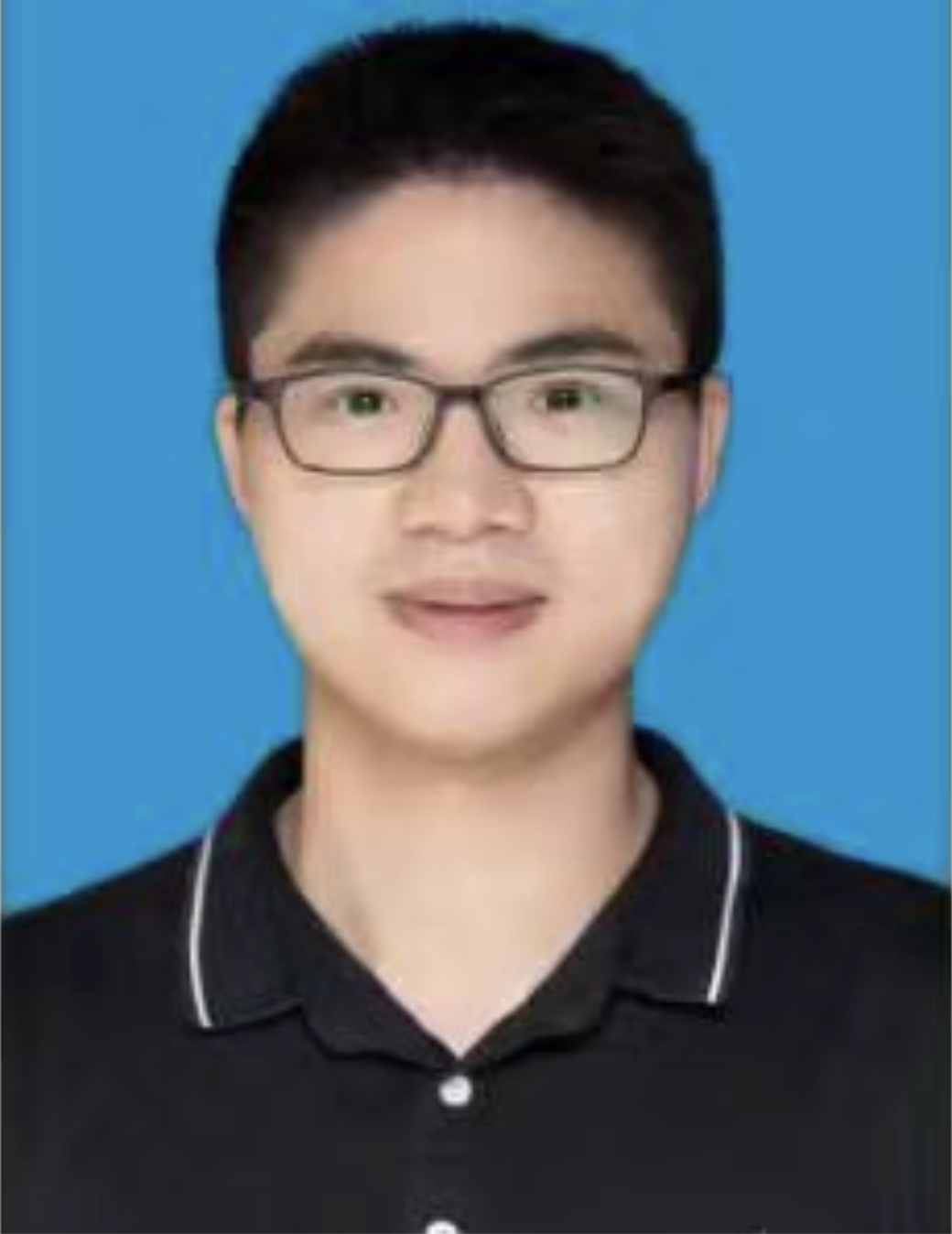}}]{JiaWen Kang} (Senior Member, IEEE) received the M.S. degree and the Ph.D. degree from the Guangdong University of Technology, China, in 2015 and 2018, respectively. He is currently a full professor at the Guangdong University of Technology. He has been a postdoc at Nanyang Technological University from 2018 to 2021, Singapore. His research interests mainly focus on blockchain, Metaverse, and AIGC in wireless communications and networking.
\end{IEEEbiography}

\begin{IEEEbiography}[{\includegraphics[width=1in,height=1.25in,clip,keepaspectratio]{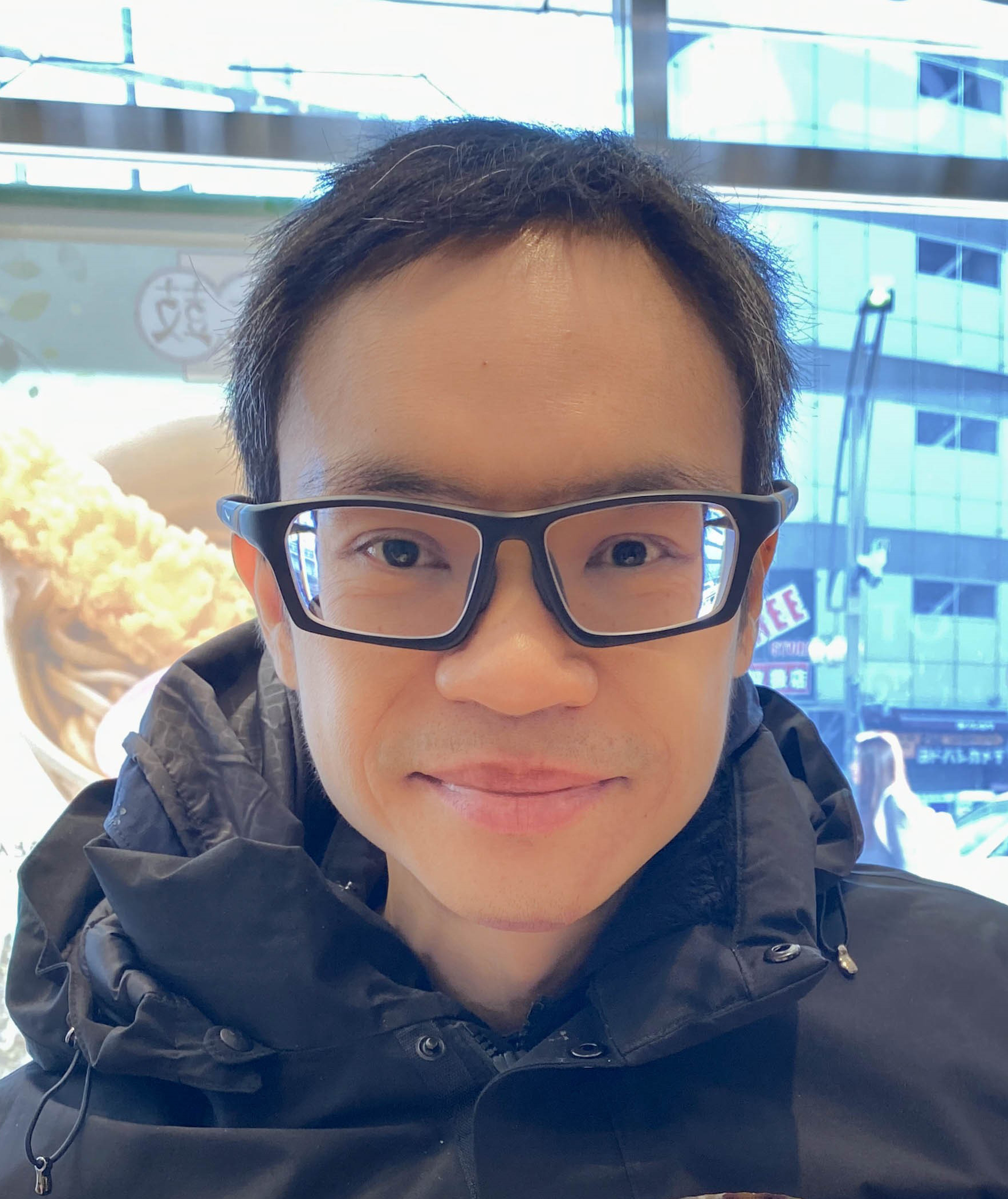}}]{Dusit Niyato } (Fellow, IEEE) is a professor in the College of Computing and Data Science, at Nanyang Technological University, Singapore. He received B.Eng. from King Mongkuts Institute of Technology Ladkrabang (KMITL), Thailand and Ph.D. in Electrical and Computer Engineering from the University of Manitoba, Canada. His research interests are in the areas of sustainability, edge intelligence, decentralized machine learning, and incentive mechanism design.
\end{IEEEbiography}

\end{document}